\documentclass[iop]{emulateapj}

\usepackage{graphicx}
\usepackage{natbib} 
\usepackage{times}
\usepackage{amsmath}
\usepackage{xcolor}
\usepackage{comment}
\usepackage{setspace}
\usepackage{longtable}
\usepackage{hyperref}

\def\apjl{{ApJ}}                
\def\apjs{ {ApJS}}               
\def\ao{ {Appl.~Opt.}}           
             
\def\aap{ {A\&A}}

\newcommand{\gtsima}{$\; \buildrel > \over \sim \;$}
\newcommand{\ltsima}{$\; \buildrel < \over \sim \;$}
\newcommand{\prosima}{$\; \buildrel \propto \over \sim \;$}
\newcommand{\gsim}{\lower.5ex\hbox{\gtsima}}
\newcommand{\lsim}{\lower.5ex\hbox{\ltsima}}
\newcommand{\simgt}{\lower.5ex\hbox{\gtsima}}
\newcommand{\simlt}{\lower.5ex\hbox{\ltsima}}
\newcommand{\simpr}{\lower.5ex\hbox{\prosima}}

\begin{document}


\title{Planetary parameters, XUV environments and mass-loss rates \\ for nearby gaseous planets with X-ray detected host-stars}

\author{Riccardo Spinelli$^{1,2}$,
Elena Gallo$^{3}$,
Francesco Haardt$^{1,2,4}$,
Andrea Caldiroli$^{5}$,
Federico Biassoni$^{1,2}$, \\
Francesco Borsa$^{2}$,
Emily Rauscher$^{3}$}

\altaffiltext{1}{Dipartimento di Scienza e Alta Tecnologia, Universit\`a degli Studi dell'Insubria, via Valleggio 11, I-22100 Como, Italy}
\altaffiltext{2}{INAF -- Osservatorio Astronomico di Brera, Via E. Bianchi 46, 23807 Merate, Italy}
\altaffiltext{3}{Department of Astronomy, University of Michigan, 1085 S University, Ann Arbor, Michigan 48109, USA}
\altaffiltext{4}{INFN, Sezione Milano-Bicocca,P.za della Scienza 3, I-20126 Milano, Italy}
\altaffiltext{5}{Fakult\"at f\"ur Mathematik, Universit\"at Wien, Oskar-Morgenstern-Platz 1, A-1090 Wien, Austria}





\begin{abstract}
We leverage Gaia DR2 parallactic distances to deliver new or revised estimates of planetary parameters and X-ray irradiation for a distance-limited ($\simlt 100$ pc) sample of 27 gaseous planets (from super-Earths to hot Jupiters) with publicly available Chandra and/or XMM observations, for which we carry out a homogeneous data reduction. 
For 20 planets with X-ray detected host stars we make use of the photoionization hydrodynamics code ATES to derive updated atmospheric mass outflow rates. The newly derived masses/radii are not consistent with the \verb|exoplanet.eu|~values for five systems; HD 149026b and WASP-38, for mass; and Au Mic b, HAT-P-20 and HAT-P-2 for radii. Notably, the lower mass implies a (Saturn-like) density of $0.86\pm 0.09$ g cm$^{-3}$) for HD 149026 b. This independent estimate is consistent with the lowest values reported in the literature. Separately, we report on the X-ray detection of GJ 9827, HD 219134 and LHS 1140 for the first time. 
The inferred stellar X-ray luminosity of LHS 1140 ($1.34^{+0.19}_{-0.21} \times 10^{26}$ erg sec$^{-1}$) implies that LHS 1140~b is the least irradiated transiting super-Earth known to orbit within the habitable zone of a nearby M-dwarf.
\end{abstract}
%



\section{Introduction}
The ongoing surge in the number of known exoplanets, including a large population of transiting systems, poses outstanding questions around their demographics, and specifically the inferred radius distribution \citep{Fulton2017}.  
One such key open question concerns the impact of X-ray and extreme UV (combined, XUV) stellar irradiation. For close-in planets, prolonged exposure to intense photo-ionizing radiation can heat and inflate planetary atmospheres, potentially leading to the removal of a substantial portion of the atmosphere's initial light element gas envelope \citep[e.g.][]{owen2013, owen2018, Wu2019,owen2019}.
Assuming that the incident stellar XUV flux is primarily converted into expansion work, the current rate of mass loss rate $\dot M$ from an irradiated atmosphere can be expected to depend directly upon the incident XUV flux and inversely upon the planetary density \citep{Watson1981,Erkaev2007}. In practice, numerical work has shown that the validity of this ``energy-limited" approximation is fairly limited, and that efficiency of this process depends (strongly) on the planet gravity, as well as the intensity of the photoionizing radiation  \citep[e.g.][]{Lammer2003,Yelle2004,Tian2005,GarciaMunoz2007,MurrayClay2009,Owen2012,Erkaev2013, Erkaev2016,Salz2016b,atesII}. \\

An accurate estimate of the XUV irradiation (at the planet surface) hinges on accurate measurements of stellar distance as well as orbital distance; the latter is often derived from stellar parameters, typically using stellar models or empirically-calibrated stellar relations. As noted by \citet{Stassun2017}, the use of stellar models may be precise but not always accurate, owing to the uncertainties in stellar evolution models, including the role of stellar rotation and its relationship with age and activity. \\

The second Gaia data release (\citealt{Gaia2018, Riello2018, Sartoretti2018, Luri2018}; hereafter GDR2) provides parallactic distance determinations for over a billion Milky Way stars. These enable more reliable estimates of stellar XUV luminosities as well as planetary parameters, and thus atmospheric mass loss rates. 
In this Paper we present new and/or revised parameters for a distance-limited sample (within 100 pc) of 27 highly irradiated gaseous systems with available X-ray observations. Specifically, we (i) perform a homogeneous X-ray spectra analysis, and provide revised X-ray luminosities--based on GDR2 distances; (ii) obtain updated planetary parameters, including density and XUV irradiation; (iii) for all the X-ray detected systems within the code convergence limits, we run the 1D photoionization hydrodynamics code ATES \citep{ates1}, and derive stable solutions for the instantaneous atmospheric mass loss rates. 

The Paper is organized as follows. In \S~\ref{sec:sample} we present the sample; \S~\ref{sec:x} describes the X-ray observations analysis and results; \S~\ref{sec:param} presents the method for estimating new planetary parameters based on the GDR2 distances; \S~\ref{sec:mdot} makes use of ATES to derive (whenever possible) new mass outflow rates. \S~\ref{sec:summary} summarizes our findings.
In a companion paper, we focus on a sub-sample of 16 out of the 27 planets considered here, and investigate how varying the intensity of the stellar XUV irradiation affects atmospheric escape \citep{atesII}.

\begin{table*}[t]
\centering
\fontsize{9}{12}\selectfont
	\caption{Planet sample; GDR2 distance (1); host star effective temperature (2); host star spectral type (3); host star radius (4); ratio between the orbital semi-major axis and the stellar radius (corrected for eccentricity unless otherwise noted) (5); orbital period (6), optical transit depth (7); radial velocity semi-amplitude (8); orbital inclination angle (9); eccentricity (10); periastron argument (11); references (12) ; $^{*}$ indicates not corrected for eccentricity.}
	\label{tab:planets_params}
	\centering
\begin{tabular}{ccccccccccccc}
	\hline
	\hline
	 & %
	\boldmath $d$        & %
	\boldmath$T_{\rm eff}$& %
	\boldmath$\rm Spectral$ &
	\boldmath$R_{\star}$   					 & %
	\boldmath$(a/R_{\star})_E$	 			 & %
	\boldmath$P$ 		 & %
	\boldmath$\Delta F$ 			 & %
	\boldmath$K$       & %
    \boldmath$i$\newline                         & %
	\boldmath$e$ 	\newline	                 &%
		\boldmath$\omega$ 	\newline &	       	
    \boldmath$Ref.$  \newline  	\\
&
     $\mathbf{[pc]}$ &
     $\mathbf{[K]}$  &
      \boldmath$\rm type$  &
        $\mathbf{[R_{\odot}]}$&
        &
      	$\mathbf{[d]}$&
      [\%]&
       $\mathbf{[m/s]}$ &
      & 
      & 
      & 
      \\
        & 
    (1) & %
    (2) & %
    (3) & %
    (4) & %
    (5) & %
    (6) & %
    (7) & %
    (8) & %
    (9) & %
    (10) &%
    (11) & 
    (12)
         \\%
	\cline{1-13}
	Host FGK & & & & & & & & & & & &  \\
	\cline{1-13}
     HAT-P-2 b (S16) &  127.77      &    6338  &    F8    &    1.73  &   8.99    & 5.633  
     & 0.522       &   983.9     &   86.72     &  0.517    &  185 & 1\\
     WASP-18 b (S16)    & 123.48 & 6462 & F6     & 1.24 & 3.48  & 0.941  & 1.04   & 1814   & 83.5   & 0.005 & -85 & 2\\
    HAT-P-20 b (S16)  & 71.03  & 4501 & K7     & 0.73 & 11.36 & 2.875  & 2.40   & 1249.5 & 86.88  & 0.017 & 343  & 3  \\
    WASP-10 b (S16)   & 141.00 & 4878 & K5V    & 0.67 & 11.90$^*$ & 3.093  & 2.48   & 543.0 & 88.81  & 0.06 & 152 & 4,5  \\
     WASP-38 b (S16)   & 136.24 & 6170 & F8     & 1.47 & 12.15 & 6.872  & 0.71   & 253.8  & 88.83  & 0.032 & -19 & 6 \\
    WASP 8 b  (S16)   & 89.96 & 5610  & G6 & 0.99&  18.20 & 8.159&   1.276& 221.1 &88.55 &0.304&274& 7 \\
   WASP-43 b (S16)	    & 86.75  & 4306 & K7V    & 0.66 & 4.97  & 0.814  & 2.52   & 551.0  & 82.11 & 0   & 90 & 3 \\
   WASP-77 A b (S16)   & 105.17 & 5556 & G8V    & 0.95 & 5.33  & 1.360  & 1.78   & 323.4  & 88.91  & 0.007 & -166 & 2\\
    HD 189733 b (S16) & 19.76  & 5015 & K2V    & 0.78 & 8.99  & 2.219  & 2.47   & 200.56 & 85.71  & 0  & 90  & 8,9 \\
   WASP-80	b  (S16)& 49.79  & 4076 & K7V    & 0.61 & 12.63 & 3.068  & 2.94   & 109.0  & 89.02  & 0.002   & 94 & 10 \\
   HD 209458 b (S16) & 48.30  & 6076 & F9V    & 1.18 & 8.78$^*$  & 3.525  & 1.49   & 84.9   & 86.71  & 0.01 & 0.0  & 8,11 \\
   HD 149026 b (S16)  & 75.86  & 6076 & G0 IV  & 1.47 & 5.98$^*$  & 2.876  & 0.27   & 37.9  & 84.50   & 0.003 &100 & 12,13\\
     WASP-29 b  & 87.60       & 4732   & K4V   &   0.81     & 12.50     & 3.923      & 0.933        &   36.0 & 89.47     &    $<0.059$    &  -           & 5,14\\
    WASP 69 b 	    & 49.96  & 4875 & K5     & 0.82 & 12.00 & 3.868  & 1.79   & 38.1   & 86.71  & 0  & 90 & 15,16  \\
    HAT-P-11 b (S16) 		&  37.77& 4757 &K4 &0.77  &  14.64 & 4.888   & 0.343& 11.6 & 88.99 & 0.265 & -162 & 17,18\\
    55 Cnc e (S16)    & 12.59  & 5306 & K0IV-V & 0.95 & 3.52  & 0.737  & 0.033  & 6.02   & 83.59  & 0.05 & 86 & 7\\
    HD 97658 b (S16)  & 21.56  & 5192 & K1V    & 0.75 & 24.2$^*$  &9.490 & 0.07   & 2.90   & 89.45  & 0.063 & -9 & 19,20\\
     WASP-107 b  & 64.74     &  4233  &   K6     &   0.73   &   18.16$^*$    & 5.721        &    2.08    &    14.1    &   89.8     &  0.06   & 40 &21,22 \\
      HD 219134 b  &  6.53     &  4787    & K3V      &  0.80 &10.48    &  3.092     &    0.0326    &  2.381      &   85.01    &   0     &  0   &  23 \\     
   GJ 9827 d    & 29.66  & 4174 & K6V    & 0.65 & 19.7 & 6.202  & 0.096  & 2.50   & 87.39  & 0   & 0 & 24 \\
     GJ 9827 b    & 29.66  & 4174 & K6V    & 0.65 & 6.62  & 1.209  & 0.059  & 2.84   & 85.73  & 0  & 0  & 24 \\
	\cline{1-13}
    Host M & & & & & & & & & & & & \\
	\cline{1-13}
	 K2-25 b & 44.96 & 3207 & M4.5 & 0.32 & 21.09 & 3.485 & 1.155 & 27.9 & 87.16 & 0.43 & 120 & 25\\ 
    GJ 436 b (S16)	    & 9.75  & 3416 & M2.5  & 0.43  & 14.54 & 2.644  & 0.68   & 17.59  & 86.86 & 0.162 & 372 & 26\\
    LHS 1140 b	& 14.99  & 3016 & M4.5  & 0.21  & 96.4  & 24.737 & 0.493  & 4.21   & 89.88  & $<0.096$ & 90 & 27\\
    AU Mic b &  9.72     & 3992     &  M1V      &   0.61   &  19.1     &  8.463      &  0.246      &  8.50 & 89.5      & 0       &  -   &  28 \\
    GJ 3470 b (S16)   & 29.42  & 3725 & M1.5  & 0.55  & 13.94 & 3.337  & 0.584  & 9.2    & 88.88  & 0.017 & 1.7 & 29\\
    GJ 1214	b  (S16)  & 14.64  & 3026 & M4.5V & 0.22 & 14.62 & 1.580  & 0.0139 & 10.9   & 89.19  & $<0.23$ & 0 & 30\\
	\hline
	\hline
	\end{tabular}
 \begin{minipage}{\linewidth}
    \vspace{10pt}
    \setstretch{0.2}
\textbf{References:} For all sources: $d$ from \citet{Gaia2018}. For all the FGK stars plus Au Mic $T_{\rm eff}$ is from \citet{Andrae2018}. For out of filter M dwarfs we use the results of \citep{Andrae2018}, assuming $T_{\rm eff}$ from \citet{Stefansoon2020} for K2-25, \citet{Vonbraun2012} for GJ 436, \citet{Spinelli2019} for LHS 1140,from \citet{Palle2020} for GJ 3470, and \citet{Aglanda2013} for GJ 1214. $R_{\star}$ is from \citet{Andrae2018} for FGK stars. For M dwarfs $R_{\star}$ are obtained from $T_{\rm eff}$ using the empirical relationships derived by \citet{Morrell2019} using GR2 data. $(a/R_{\star})_E$, $P$, $\Delta F$, $K$, $i$, $e$ and $\omega$ are from the following references: (1)~\citet{Bakos2010}; (2)~\citet{Cortes-Zuleta2020}; (3)~\citet{Esposito2017}; (4)~\citet{Barros2013}; (5)\citet{bonomo2017}; (6)~\citet{Simpson2011}; (7)~\citet{Bourrier2018}; (8)~\citet{Southworth2010}; (9)~\citet{Cegla2016}; (10)~\citet{Triaud2015}; (11)~\citet{Rosenthal2021}; (12)~\citet{Stevenson2012}; (13)~\citet{Kutson2014}; (14)~\citet{Saha2021};	(15)~\citet{Anderson2014}; (16)~ \citet{Casasayas-Barris2017};  (17)~\citet{Huber2017}; (18)~\citet{Yee2018}; (19)~\citet{Ellis2021}; (20)~\citet{Dragomir2013}; (21)~\citet{Piaulet2021}; (22)~\citet{Anderson2017};  (23)~\citet{Gillon2017};(24)~\citet{Rodriguez2018}; (25)~\citet{Stefansoon2020};  (26)~\citet{Lanotte2014};(27)~\citet{LilloBox2020};(28)~\citet{Martioli2021}; (29)~\citet{Biddle2014}; (30)~\citet{Aglanda2013}
 \end{minipage}
\end{table*}

\newpage
\section{Sample}
\label{sec:sample}
We draw our targets from the \verb|exoplanet.eu|\footnote{\url{http://exoplanet.eu/}} database, selecting planetary systems with (i) both transit and radial velocity data; (ii) (pre-Gaia DR2) distances within $\sim$100 pc; (iii) average orbital distances within 0.1 A.U., and; (iv) planetary radii in excess of 1.5 $R_{\Earth}$ (or 0.134 $R_J$), above which a primordial (H/He-rich) atmosphere may be retained \citep{Lopez2014}. 
Out of these, we sub-select for host stars with publicly available X-ray spectroscopic data, either with XMM-Newton or the Chandra X-ray Observatory, yielding a final sample of 26 planetary systems comprising 27 planets. A complete list of planetary parameters can be found in Table \ref{tab:planets_params}, where the planets are listed in order of decreasing mass.  

Below, we provide a brief description of each system, with an emphasis on recent constraints or predictions concerning atmospheric escape and its detectability; following \citet{Kopparapu2018}, we divide planets into rocky (0.5 -- 1.0 R$_{\Earth}$), super-Earths (1.0 --
1.75 R$_{\Earth}$), sub-Neptunes (1.75 -- 3.5 R$_{\Earth}$), sub-Jovians (3.5 -- 6.0 R$_{\Earth}$) and Jovians
(6 -- 14.3 R$_{\Earth}$) based on their planet sizes, and ``hot'', ``warm'' and ``cold'' based
on the incident stellar flux (see table 3 in \citealt{Kopparapu2018} for details). 

\subsection{Jovians}
Together with WASP-80b \citep{Triaud2013} and WASP-10 \citep{Christian2009}, WASP-43b \citep{Hellier2011} represents one of the rare instances of a transiting hot Jupiter around a late K/early-M star. This makes them promising targets for detecting and probing atmospheric escape \citep{Oklopcic2019}, albeit the recent non-detection of metastable helium in the transit spectrum of WASP-80b poses considerable theoretical challenges \citep{fossati22}.

WASP-69b \citep{Anderson2014} is a hot Jupiter orbiting an active K-type star. \citet{Nortmann2018} reports on the detection of excess helium absorption during transit; the complex line shape during and post transit is consistent with an out-flowing cometary atmosphere.

HD 149026b \citep{Sato2005} is a Jovian planet with mass close to Saturn’s, orbiting a G star. Several studies \citep{Sato2005, Charbonneau2006, Winn2008, Nutzman2009} conclude that the planet is significantly denser than Saturn, which would be consistent with a high metallicity. The presence of metals in its atmosphere has also been proposed to explain the inferred high density of HAT-P-20~b \citep{Deming2015} compared to other systems with similar masses \citep{Bakos2011}. 

HD 189733b \citep{Bouchy2005} and HD 209458b \citep{Charbonneau2000, Henry2000} are among the closest (and hence best studied) transiting hot Jovian planets. The detection of Lyman-$\alpha$ absorption in transit in these systems provided the first observational evidence for an escaping atmosphere  \citep{Vidal-Madjar2003, Lecavelier2010}. 
\citet{Salz2018} and \citet{Alonso2019} have also reported the detection of metastable helium absorption in the upper atmosphere of HD 189733b and HD 209458b. To explain the different helium absorption depths measured in different transits, \citet{Zhang2022} proposed stellar XUV variability.

WASP-77Ab \citep{Maxted2013} is a transiting hot Jovian orbiting around a moderately bright G-type star (the member of a visual binary). WASP-18b \citep{Hellier2009} is an extremely close-in hot Jupiter, with 10 M$_{J}$ and an orbital period of 0.94 days. In spite of the extreme orbital parameters, this system is remarkably inactive \citep{miller12}. Similarly, the large mass ($8$ M$_{\rm J}$) and relatively short orbital period ($\sim$5.63 d) of HAT-P-20~b \citep{Bakos2007} make it a promising target for planet-star interactions studies.

WASP 107b \citep{Anderson2017} is an extremely low density Jovian planet (0.13 g cm$^{-3}$, \citealt{Piaulet2021}) and associated with the first detection of metastable helium during transit \citep{Spake2018}. WASP-38b \citep{Barros2011} is a Jovian planet that surprisingly moves in prograde orbits well aligned, despite of the orbital eccentricity \citep{brown2012}.

HAT-P-2b \citep{Bakos2010} and WASP-8b \citep{Queloz2010} are two hot Jupiters in eccentric orbits. The orbit of WASP-8b is misaligned with the stellar rotation axis, and moving in a retrograde direction. WASP-29b \citep{Hellier2010} is a low-density Jovian planet for which the reported Lyman-$\alpha$ upper limit argues against the presence of an escaping hydrogen atmosphere \citep{DosSantos2021}.

\subsection{Sub-Jovians}
GJ 436b \citep{Butler2004}, GJ 3470b \citep{Bonfilis2012}, HAT-P-11b \citep{Bakos2010} and  K2-25b \citep{Mann2016} are the four closest, transiting sub-Jovians planets. As noted by \citet{King2018} only one other transiting planet within 100 pc (HD 3167c) has a radius between 3 and 5 $R_{\Earth}$. GJ 436b is the most illustrative example for the presence of a planetary tail of material, inferred trough Lyman-$\alpha$ observations \citep{Kulow2014, Ehrenreich2015, Lavie2017}. \citet{Nortmann2018} provided only an upper limit for helium signals in the atmospheres of GJ 436b. Lyman-$\alpha$ absorption during GJ 3470 b transits is observed by \citet{Bourrier2018}, while \citet{Mansfield2018} report the detection of helium in the atmosphere of HAT-P-11 b. K2-25b orbits a relatively young ($\sim$727 Myr) M4.5 host with a period of 3.48 days. The planet is similar to GJ 436b, although \citet{Rockcliffe2021} report on a Lyman-alpha non-detection during transit. In addition, the combination of the inferred high-density and young age for this system challenges prevailing theories of planet formation \citep{Stefansoon2020}.

Au Mic b \citep{Plavchan2020} is a sub-Jovian planet just above the radius gap, orbiting a young (22 Myr) fastly rotating ($\sim$4.85 d) pre-main sequence M dwarf. It represents a rare target for studying atmospheric loss during early evolutionary stages, when the host star is likely very active and has a much higher XUV bolometric ratio than after the first 100 Myr \citep[e.g.,][]{Vilhu1984, Vilhu1987, Wright2011}. 

\subsection{Sub-Neptunes}
HD 97658b \citep{Howard2011} is a moderately irradiated sub-Neptune whose low density (3.9 g cm$^{-3}$, \citealt{VanGrootel2014}) is compatible with a massive steam envelope that could be dissociated by stellar irradiation at high altitudes. Nevertheless, the non-detection of neither Lyman-$\alpha$ nor helium during transit suggests that the system may lack an extended, evaporating hydrogen atmosphere \citep{Bourrier2017, Kasper2020}. 

55~Cnc~e \citep{McArthur2004} is an ultra-short period planet (P$<$1 day) whose radius sits right at the upper limit of the so-called small planet ``radius gap" \citep{Fulton2017}, i.e., between 1.5 and 2 R$_{\Earth}$. The absence of helium and Lyman-$\alpha$ absorption during transit \citep{Ehrenreich2012,Zhang2021} suggests that the planet has probably already lost its envelope.

GJ 9827b,d \citep{Rodriguez2018} are two planets orbiting a bright K6 star with planetary radii that span across the radius gap. In particular, GJ 9827b's radius falls between that of super-Earths and sub-Neptunes \citep{Fulton2017}. \citet{Carleo2021} and \citet{Kasper2020} reported a non-detection of helium absorption for GJ 9827 b and GJ 9827 d.

GJ 1214b \citep{Charbonneau2009} is a sub-Neptune orbiting an M-dwarf; it is best known for the claim that high-altitude clouds would obscure the deeper atmospheric layers \citep{Kreidberg2014}.  Recently, \citet{Orell2022} reported a significant ($>$ 4$\sigma$) detection of metastable helium in this system during one transit observed with the CARMENES spectrograph, in contrast with previous non-detections \citep{Kasper2020,Petit2020}. New observations are expected to eventually confirm the presence of an extended atmosphere. 

\subsection{Super-Earths}

LHS 1140b \citep{Dittmann2017} is a super-Earth orbiting an M-dwarf. LHS 1140 b is located within the so-called habitable zone, where temperature and pressure would be suitable for the presence of liquid water on the planetary surface \citep[e.g.][]{Kasting1993, Kopparapu2013}.  As such, this system represents a key target for future astrobiological studies \citep{Spinelli2019, Wunderlich2021, Edwards2021}.  

HD 219134 b \citep{Motalebi2015} is the closest transiting exoplanet and the innermost of a planetary system with other four planets in tight orbit ($<$0.4 AU) and one distant Jovian planet (3 AU), with an architecture roughly similar to the solar system’s one \citep{Vogt2015}. \citet{Folsom2018} derive the the stellar wind mass-loss rate of the star of the system (10$^{-14}$ M$_{\Sun}$ yr$^{-1}$, half the solar value) using HST/STIS observations of the Lyman-$\alpha$ line.

\begin{table}
\centering
\renewcommand{\arraystretch}{1.0}
\fontsize{7}{12}\selectfont
\caption{X-ray observation log; instruments (1) (C for Chandra, X for XMM), observation ID (2), observation date (3), exposure time (4), Principal Investigator (5).}
	\begin{tabular}{
cccccc}
    	\hline
    	\hline
    	       &%
    	Instr. &%
    	ObsID  &%
    	Date   &%
        Exp. time&%
        PI \\ 
        	         &%
        	         &%
        	         &%
        	         &%
        [ks]	 &%
    				\\%
            &%
        (1) &%
        (2) &%
        (3) &%
        (4) &%
        (5) \\
    	\cline{1-6}
    	Host FGK		  & &  & 	& & \\
    	\cline{1-6}
        HAT-P-2 & C & 15707 & 2013-11-16 & 10 &Salz \\
        WASP-18      & C & 14566 & 2013-02-26 & 85  & Pillitteri     \\%
        HAT-P-20     & C & 15711     & 2013-11-24 & 10 & Salz         \\%
        WASP-10      & C & 15710	   & 2013-11-15 & 10 & Salz         \\%
        WASP-38      & C & 15708      & 2014-01-18 & 10 & Salz         \\%
        WASP 8       & C & 15712     & 2013-10-23 & 10 & Salz         \\%
        WASP-43 & X &	0694550101 & 2012-05-11 & 18 & Grosso       \\%
        WASP-77A      & C & 15709      & 2013-11-09 & 10 & Salz         \\%
        WASP80	          & X&0744940101 & 2014-05-15(A) & 19& Salz\\
        & X & 0764100801 & 2015-05-13 (B) & 33 & Wheatley    \\
        HD 209458   & X  & 0130920101 & 2000-11-29(A) & 19 & Jansen \\%
                & X & 0148200101 & 2003-06-11(B) & 23 & Bertaux\\
        & X & 0404450101 & 2006-11-14(C) & 34 & Wheatley     \\
        & C & 16667 & 2016-06-17(D) &29& Czesla \\
        HD 149026    & X & 0763460301 & 2015-08-14 & 20 & Salz         \\%
        WASP 29  &  X  &  0804790201         &      2017-05-14       &  8  & Sanz-Forcada    \\
                WASP-69 	  & X & 0783560201 & 2016-10-21 & 31 & Salz         \\%
        HAT-P-11	  & X & 0764100701 & 2015-05-19(A) & 32 & Wheatley	    \\
        	  & C& 16669 & 2015-11-14(B) & 8 & Miller	    \\%
        55 Cnc      & X & 0551020801 & 2009-04-11(A) & 14 & Schmitt      \\%
                    & C& 14401 & 2012-03-07(B) & 11 & Wheatley  \\
                    & C & 14402 & 2012-04-05(C) & 20 & Wheatley  \\
         HD97658 & X &	0764100601 & 2015-06-04(A) & 34 & Wheatley     \\
         & C & 16668 &2015-10-17(B)  & 13& Miller      \\
        & C & 18724& 2015-12-11(C)&  20 &  Wheatley \\
         & C & 18725&  2016-03-05(D)& 20 &  Wheatley  \\%
        WASP 107  & X   & 0830190901   &    2018-06-22       &     63        &   Shartel     \\
         HD 219134  & X& 0784920201  &   2016-06-13        &   38          &  Wheatley       \\
        GJ 9827 	  & X & 0821670101 & 2018-05-27(A) & 22 & Drake        \\
        & X & 0821670201 & 2018-06-23(B) & 17 &  Drake \\
          & X & 0821670301 & 2018-11-27(C) & 11  & Drake\\
    	\cline{1-6}
        Host M & & & & &  \\
    	\cline{1-6}
    	 K2-25 & X & 	0782061001 & 2017-03-07 & 19 & Agueros \\
        GJ 436 	      & X & 0556560101 & 2008-12-10(A) & 33 & Wheatley\\
        &C& 14459& 2013-02-16(B)& 19 & Ehrenreich \\
                &C& 15537& 2013-04-18(C)& 19& Ehrenreich\\
        &C& 15536& 2013-06-18(D)& 20& Ehrenreich\\
        &C& 15642& 2014-06-23(E)& 19& Ehrenreich\\
       &C& 17322& 2015-06-25(F)& 9& France\\
        &C& 17321& 2015-06-26(G)&20& France\\
       &X  &0764100501 & 2015-11-21(H) & 27 & Wheatley \\
       LHS 1140 	  & X & 0822600101 & 2018-12-21 & 64 & Dittmann    
        \\
        Au Mic & X & 0111420101 & 2000-10-13 (A) & 56&Brinkman \\
        & X & 0822740301 & 2018-10-10 (B) & 134&  Kowalski\\
        & X& 0822740401 & 2018-10-12 (C)& 137&  Kowalski\\
        & X& 0822740501 & 2018-10-14 (D) & 139&  Kowalski\\
        & X& 0822740601 & 2018-10-16 (E) & 70 &  Kowalski\\
        GJ 3470        & X & 0763460201 & 2015-04-15 & 18 & Salz         \\%
        GJ 1214	      & X & 0724380101 & 2013-09-27(A) & 36 & Sairam       \\
         & C & 15725& 2014-06-07(B) &31  &Brown      \\%

    	\hline
    	\hline
	\end{tabular}
\label{tab:x2}
\end{table}

\begin{table}
\centering
\renewcommand{\arraystretch}{1.2}
\fontsize{7}{11}\selectfont
\caption{Observation log for HD 189733 (S16); instruments (1) (C for Chandra, X for XMM), observation ID (2), date (3), exposure time (4), Principal Investigator (5).}
	\begin{tabular}{ ccccc} 
    	
    	\hline
    	\hline
    	ObsID  &%
    	Instr. &
    	Date   &%
        Exp. time&%
        PI \\ 
        & 
        & 
        [ks]  & 
        \\
        (1) &%
        (2) &
        (2) &%
        (3) &%
        (4) \\
    	\cline{1-5}
         X   &  050607201  & 2007-04-18(A) & 55 &Wheatley \\
         X   &	 0600970201 &		2009-05-18(B)& 37& Wolk\\
          X   &	 0672390201 &		2011-04-30(C)& 39& Pillitteri\\
          C& 12340& 2011-07-05(D) &19 & Poppenhaeger\\
          C& 12343& 2011-07-12(E)& 20& Poppenhaeger\\
        C& 12344& 2011-07-16(F)& 18 &Poppenhaeger\\
        C& 12345& 2011-07-18(G) & 20 &Poppenhaeger\\
        C& 12341& 2011-07-21(H)& 20& Poppenhaeger \\  
         C& 12342& 2011-07-23(I)& 20& Poppenhaeger\\
        X   &  0690890201 &  2012-05-07(J)	      &    62&  Pillitteri	    \\
         X   &	 0692290201 &		2013-05-09(K)& 39& Wheatley\\
          X   &	 0692290301 &		2013-11-03(L)& 37& Wheatley\\
         X  &	 0692290401 &	2013-11-21(M) &42 &Wheatley \\
          X  &   0744980201 & 2014-04-05(N) & 48  &Wheatley \\
            X   &	 0744980301 &2014-05-02(O)& 34&Wheatley \\
           X   &	 0744980401 &	2014-05-13(P)&40 &Wheatley \\
            X   &	 0744980501 &2014-05-15(Q)& 32& Wheatley\\
          X   &	 0744980601 &2014-05-17(R)& 32& Wheatley\\    
        X   &	 0744980801 &		2014-10-17(S)& 37& Wheatley\\
                  X   &	 0744980901 &2014-10-19(T)& 34& Wheatley\\
        X   &	 0744981001 &	2014-10-22(U)&40 & Wheatley\\
        X   &	 0744981101 &		2014-10-24(V)& 39& Wheatley\\
        X   &	 0744981301 &2014-11-08(W)& 35& Wheatley\\
            X  &	 0744981201 &	2014-11-11(X)& 44& Wheatley\\
         X   &	 07449814901&2014-11-13(Y)& 32& Wheatley \\    
          X   &	 0744980701 &		2014-11-15(Z)& 39& Wheatley\\

          X  &   0748391401 &  2015-04-03(AA) &47 &Schartel \\
          X  &	 0744981501 &	2015-04-13(AB) & 45& Wheatley \\
                  X   &  0744981601 &		2015-04-17(AC)& 41&Wheatley \\
                       
          X   &	 0744981701 &		2015-04-19(AD)& 38& Wheatley\\
          X  &	 0748391501 &	2015-04-23(AE) & 44& Schartel\\
    	\hline
	\end{tabular}
\label{tab:x3}
\end{table}

\begin{table*}
\centering
\caption{Best-fit X-ray spectral models; Hydrogen column density (1); best fit model temperature (2, 3 and 4); best fit emission measure (5,6 and 7) (10$^{50}$ cm$^{-3}$); unabsorbed flux at Earth, in 10$^{-14}$ erg s$^{-1}$ cm$^{-2}$ (8).}
\label{tab:flux}
\fontsize{9}{14}\selectfont
\renewcommand*{\arraystretch}{1.0}
\begin{tabular}{
cccccccccc}
    	\hline
    	\hline
    	                       &                              & %
    	\boldmath $n_H$       & 
    	\boldmath $KT_{1}$   & 
        \boldmath $KT_{2}$   &  %
        \boldmath $KT_{3}$    & %
    	\boldmath $EM_1$    & %
    	\boldmath $EM_2$   &  %
    	\boldmath $EM_3$   & %
        \boldmath $F_X$ 			            \\%
   & &
        \newline $\mathbf{[10^{18} cm^{-2}]}$ & %
        \newline $\mathbf{[keV]}$             & %
        \newline $\mathbf{[keV]}$ 		    &
        \newline $\mathbf{[keV]}$ 		    &
        \newline						$\mathbf{[cm^{-3}]}$		&
        \newline					$\mathbf{[cm^{-3}]}$		    &
        \newline					$\mathbf{[cm^{-3}]}$		    & \newline(a)	\\
           &
            & %
        (1) & %
        (2) & %
        (3) & %
        (4) & %
        (5) & %
        (6) & %
        (7) & %
        (8)  \\
    	\cline{1-10}
    	Host FGK & & & & & & & & & \\
    	\cline{1-10}
         HAT-P-2    &    & 39.4 &  0.32$^{+0.04}_{-0.03}$ &  -         &    -         & 61.96 $^{+11.55}_{-10.47}$   & - & - & 5.50$^{+0.74}_{-0.79}$   \\
         WASP 18    &    & 38.11      &   0.30   &   -   &   -   & 1.15 & - & - & $<0.10$    \\
         HAT-P-20   &    & 22.0   & 0.41$^{+0.13}_{-0.01}$ & -  & -  & 5.44$^{+1.85}_{-1.48}$   & -      & - & 1.78$^{+0.28}_{-0.35}$         \\
         WASP 10    &    & 43.5 & 0.30   & - & -    & 1.34  & -  & - & $<0.90$ \\
         WASP 38    &    & 42.04&   0.30  &  -    &  -    & 7.71  & - & -&    $<0.56$                \\
         WASP 8     &    & 27.8   & 0.29$^{+0.04}_{-0.03}$ & -  &-& 20.57$^{+4.99}_{-4.19}$   & -       & - & 3.50$^{+0.61}_{-0.71}$    \\
         WASP 43    &    & 26.8 & 1.08$^{+0.38}_{-0.28}$ & 0.18$^{+0.06}_{-0.04}$ & - &  1.39$^{+1.29}_{-0.75}$ & 2.52$^{+1.05}_{-0.87}$& - &0.75$^{+0.12}_{-0.20}$   \\
         WASP 77    &    & 32.6   & 0.75$\pm$0.11 & - & -   & 6.60$^{+1.58}_{-1.36}$    & -   & -    & 1.24$^{+0.16}_{-0.35}$  \\
         HD 189733  & A  & 6.1 &      0.67$^{+0.03}_{-0.02}$& 0.20$\pm$0.01  &   -  & 2.88$^{+0.19}_{-0.21}$   & 4.91$^{+0.17}_{-0.24}$& - & 33.46$\pm$0.74 \\
                    & B  & 6.1  &  0.74$^{+0.02}_{-0.01}$         &    0.19$\pm$0.01         & -  & 6.15$\pm$0.18     & 6.14$\pm$0.18 & - & 48.84$^{+0.75}_{-0.66}$    \\
                    & C  & 6.1  &    0.79$\pm$0.02         &    0.22$\pm$0.01        & -  & 4.17$^{+0.12}_{-0.18}$    & 6.67$\pm$0.25 & - & 46.86$^{+0.67}_{-0.60}$     \\
                    & D  & 6.1  &   0.79$^{+0.06}_{-0.04}$        &     0.22$^{+0.03}_{-0.02}$           & -  &  3.41$^{+0.29}_{-0.37}$   & 4.69$^{+0.45}_{-0.43}$ & - & 33.59$^{+1.00}_{-1.37}$ \\
                    & E  &  6.1 &   0.71$\pm$0.03        &   0.22$\pm$0.02      & -  &  3.35$^{+0.29}_{-0.30}$     & 5.17$^{+0.46}_{-0.44}$ & - &35.17$^{+1.18}_{-1.51}$  \\        
                    & F  &  6.1 &   1.00$^{+0.14}_{-0.12}$       &     0.28$\pm$0.02         & -  & 1.91$^{+0.20}_{-0.24}$      &  6.02$^{+0.52}_{-0.41}$ & - &29.72$^{+1.24}_{-1.15}$   \\      
                    & G  &  6.1 &     1.40$^{+0.25}_{-0.18}$      &    0.29$\pm$0.01           & -  &   2.58$^{+0.39}_{-0.39}$   & 7.81$\pm$0.39 & - & 34.68$^{+1.53}_{-1.08}$     \\            
                    & H  & 6.1  &   1.04$^{+0.10}_{-0.14}$         &   0.28$^{+0.01}_{-0.02}$          & -  &  1.62$\pm$0.18    & 6.77$^{+0.39}_{-0.49}$ & - & 30.64$\pm$1.12    \\
                    & I  & 6.1  &    0.75$^{+0.09}_{-0.06}$         &  0.23$^{+0.03}_{-0.02}$           & -  & 2.22$^{+0.36}_{-0.46}$     & 4.75$^{+0.58}_{-0.43}$ & - &27.84$^{+0.82}_{-1.00}$     \\         
                    & J  & 6.1    & 0.75$^{+0.01}_{-0.02}$& 0.20$\pm$0.01 &  -   & 3.28$^{+0.15}_{-0.11}$      & 5.32$\pm$0.13  & -    &  37.02$^{+0.60}_{-0.36}$\\
                    & K  & 6.1  &    0.71$^{+0.03}_{-0.02}$       &   0.20$\pm$0.01      & -  &  3.48$^{+0.14}_{-0.22}$     & 5.67$\pm$0.17 & - & 39.36$^{+0.51}_{-0.57}$    \\         
                    & L  & 6.1  &   0.75$^{+0.04}_{-0.02}$         &      0.21$\pm$0.01        & -  &  3.38$^{+0.20}_{-0.31}$   & 5.81$^{+0.40}_{-0.22}$ & - & 39.56$^{+0.59}_{-0.76}$   \\
                    & M  & 6.1  &  1.18$^{+0.06}_{-0.03}$         &    0.27$\pm$0.01  &  - & 3.25$^{+0.17}_{-0.16}$ &  7.91$^{+0.18}_{-0.17}$   &   - &  41.31$^{+0.60}_{-0.51}$   \\         
                    & N  & 6.1  &    0.70$^{+0.02}_{-0.03}$      &   0.19$\pm$0.01    & -  & 3.01$^{+0.20}_{-0.19}$ & 5.41$\pm$0.21   &  -  & 35.80$^{+0.82}_{-0.97}$    \\
                    & O  & 6.1  &    0.71$^{+0.05}_{-0.02}$        &     0.20$\pm$0.01       & -  & 3.68$\pm$0.18    & 6.12$\pm$0.21 & - &42.26 $^{+0.87}_{-0.80}$    \\         
                    & P  & 6.1  &   1.12$^{+0.08}_{-0.14}$        &   0.27$\pm$0.01            &  - &  2.61$\pm$0.15   & 8.16$\pm$0.19 & - &40.74$^{+0.61}_{-0.64}$     \\
                    & Q  & 6.1  &   0.75$^{+0.02}_{-0.01}$         &      0.20$\pm$0.01         & -  &  5.01$^{+0.20}_{-0.19}$   & 6.15$\pm$0.22 & - &49.40$^{+0.81}_{-0.70}$\\         
                    & R  & 6.1  &   0.88$^{+0.02}_{-0.06}$        &   0.23$^{+0.01}_{-0.02}$          & -  & 3.78$^{+0.34}_{-0.18}$     & 7.21$^{+0.28}_{-0.54}$ & - &45.69$^{+0.77}_{-0.75}$   \\         
                    & S  & 6.1  &    0.75$^{+0.05}_{-0.02}$         &   0.20$\pm$0.01         &  - & 4.15$^{+0.16}_{-0.43}$     & 5.74$^{+0.48}_{-0.19}$  & - & 43.37$^{+0.69}_{-0.74}$    \\      
                    & T  & 6.1  &   1.25$^{+0.04}_{-0.09}$        &     0.28$\pm$0.01        & -  & 3.84$\pm$0.19    & 8.13$\pm$0.19 & - & 43.51$^{+0.68}_{-0.75}$     \\
                    & U  & 6.1 &      0.70$\pm$0.02     &    0.20$\pm$0.01         &  - &  4.71$^{+0.19}_{-0.18}$  & 7.02$\pm$0.21 & - & 51.14$^{+0.75}_{-0.60}$     \\         
                    & V  & 6.1  &  0.79$^{+0.04}_{-0.02}$         &   0.21$\pm$0.01          & -  &  4.27$^{+0.15}_{-0.26}$    & 5.80$^{+0.33}_{-0.18}$ & - & 43.87$^{+0.70}_{-0.44}$     \\
                    & W  & 6.1  &    1.06$^{+0.08}_{-0.12}$  & 0.28$\pm$0.01 & -  & 2.39$^{+0.16}_{-0.15}$  & 6.61$\pm$0.21  & - & 33.16$^{+0.53}_{-0.55}$  \\         
                    & X  & 6.1  &   0.94$^{+0.03}_{-0.06}$        &    0.24$\pm$0.01  & - &  3.50$^{+0.16}_{-0.15}$ &   7.02$^{+0.28}_{-0.47}$   & -   & 42.75$^{+0.58}_{-0.75}$    \\
                    & Y  & 6.1  &   0.71$^{+0.03}_{-0.02}$        &   0.20$\pm$0.01             & -  &  4.63$^{+0.20}_{-0.33}$   & 6.54$^{+0.35}_{-0.23}$& - &48.97$^{+0.76}_{-0.70}$     \\        
                    & Z  & 6.1  &   0.75$\pm$0.02        &    0.21$\pm$0.01        & -  &  4.57$^{+0.21}_{-0.19}$   & 6.86$\pm$0.24 & - & 49.86$^{+0.72}_{-0.77}$     \\
                    & AA & 6.1  &     0.78$\pm$0.02      &   0.22$^{+0.01}_{-0.02}$   & -  & 5.24$^{+0.19}_{-0.22}$  &   7.80$^{+0.28}_{-0.25}$   & -   &  56.75$^{+0.57}_{-0.67}$   \\
                         & AB & 6.1  &  0.75$^{+0.04}_{-0.02}$         &  0.23$\pm$0.01    & -  & 3.62$^{+0.20}_{-0.36}$ &  7.19$^{+0.51}_{-0.26}$   & -   &  46.04$^{+0.51}_{-0.65}$  \\
                    & AC & 6.1  &   0.83$\pm$0.01        &   0.24$\pm$0.01  & -  &  4.14$^{+0.27}_{-0.19}$   & 7.34$^{+0.29}_{-0.28}$  & - &  48.66$^{+0.56}_{-0.64}$   \\
                    & AD & 6.1  &    0.74$\pm$0.02       &    0.20$\pm$0.01        & -  &  4.77$^{+0.18}_{-0.25}$   & 6.97$^{+0.28}_{-0.21}$ & - & 51.29$\pm$0.76    \\
                    & AE & 6.1  &    0.74$^{+0.02}_{-0.01}$       &  0.20$\pm$0.01    & - & 5.00$\pm$0.18 & 7.27$^{+0.21}_{-0.20}$      &  -  &  53.46$^{+0.72}_{-0.76}$   \\
\end{tabular}
\end{table*}
\begin{table*}
\centering
\fontsize{9}{14}\selectfont
\renewcommand*{\arraystretch}{1.0}
\begin{tabular}{cccccccccc 
     }
        WASP-80     & A  & 15.4 & 0.16$\pm$0.03 & 0.73$\pm$0.09 & -  & 1.75$^{+0.56}_{-0.38}$    & 0.97$\pm$0.18 & -   & 1.77$^{+0.16}_{-0.28}$     \\  	  
                    & B  & 15.4 & 0.19$\pm$0.02 & 0.84$^{+0.12}_{-0.10}$ & - & 2.04$^{+0.24}_{-0.23}$    & 0.67$\pm$0.15   & - & 1.70$^{+0.11}_{-0.19}$     \\ 
        HD 209458   & A  & 14.9   &   0.93  & - &  -& 0.45 & -  &-& $<0.35$ \\
                    & C  &  14.9 &  0.93 & - & - & 0.23 & - & - & $<0.18$ \\
                   & D  &  14.9  & 0.93    & - & -  & 2.75$^{+1.23}_{-2.75}$ &-  & - & 2.25$^{+1.12}_{-2.25}$    \\
        HD 149026   &    & 23.5   & 0.78$^{+0.16}_{-0.11}$ & 0.20$^{+0.12}_{-0.17}$ & - & 1.36$^{+0.60}_{-0.59}$   & 1.06$^{+0.77}_{-0.63}$  & -  & 0.75$^{+0.04}_{-0.21}$   \\
        WASP 29     &    & 27.03 & 0.30  &      -     &     -        & 2.56 &  -& - & $<0.47$ \\
        WASP 69	    &    &  15.5 & 0.82$^{+0.05}_{-0.04}$  & 0.21$\pm$0.02 & - & 3.51$^{+0.36}_{-0.35}$    & 5.49$^{+0.48}_{-0.44}$   & - & 6.02$^{+0.27}_{-0.20}$    \\
        HAT-P-11    & A  &  11.7   & 0.18$\pm$0.02 & 0.87$^{+0.08}_{-0.07}$ & - & 2.09$^{+0.22}_{-0.20}$    & 8.79$\pm$0.14   &- & 3.25$^{+0.17}_{-0.28}$    \\
                    & B  &  11.7   & 0.67$^{+0.11}_{-0.09}$ & - & - & 5.39$^{+0.77}_{-0.69}$ & - & - & 7.84$^{+0.84}_{-0.72}$\\
        55 Cnc      & A  & 3.9  & 0.62$^{+0.07}_{-0.08}$ & 0.11$\pm$0.01 &- & 0.15$\pm$0.02   & 1.11$^{+0.43}_{-0.29}$  & -  & 8.65$^{+0.71}_{-1.96}$    \\
                   & B  & 3.9  &     0.18 & -& -& 0.36 & -& -& 1.89  \\
                    & C  & 3.9  & 0.18$\pm$0.03 & - & - & 0.10$^{+0.08}_{-0.04}$ & - & - & 0.77$^{+0.32}_{-0.19}$    \\
        HD 97658    & A  & 6.7  & 0.24$\pm$0.01 & 0.026$^{+0.006}_{-0.005}$& - &0.55$\pm$0.05    & 415$^{+895}_{-313}$ & - & 3.20$^{+0.10}_{-1.39}$   \\ 
                    & B  & 6.7  & 0.22$^{+0.05}_{-0.04}$ & - & - & 0.61$^{+0.39}_{-0.23}$ &- &- & 1.73$^{+0.61}_{-0.75}$\\
                    & C  & 6.7  & 0.23$^{+0.07}_{-0.05}$ & - & - & 0.34$^{+0.31}_{-0.16}$ & - & - & 0.98$^{+0.35}_{-0.51}$\\ 
                    & D  & 6.7& 0.22& - & - &  0.50 & - & - & $<1.32$ \\ 
         WASP-107   &    & 20.0 &  0.10$\pm$0.03 & 0.31$^{+0.06}_{-0.05}$ &  1.39$^{+0.28}_{-0.18}$  & 2.48 $^{+3.92}_{-0.97}$   & 1.59$^{+0.37}_{-0.38}$  & 1.21$^{+0.43}_{-0.33}$   &    1.42 $^{+0.01}_{-0.02}$  \\
        HD 219134   &    & 2.0  &  0.10$\pm$0.01   & 0.26$\pm$0.03  & -   & 0.63  $^{+0.37}_{-0.11}$      &  0.16$^{+0.05}_{-0.04}$   &  -  &   18.07$^{+0.40}_{-0.89}$  \\
        GJ 9827     & A  & 9.2  & 1.36$^{+0.80}_{-0.69}$ & 0.25$^{+0.05}_{-0.13}$ & - & 0.20$^{+0.18}_{-0.11}$ & 0.25$^{+0.08}_{-0.07}$ & -   &  0.68$^{+0.13}_{-0.12}$       \\
                    & B  & 9.2 & 0.25$^{+0.03}_{-0.02}$ & - & -& 0.41$^{+0.09}_{-0.08}$  & - & - & 0.65$^{+0.10}_{-0.15}$\\
                    & C  & 9.2&0.31$^{+0.08}_{-0.05}$& - & - & 0.25$^{+0.08}_{-0.07}$ & -  &  -    &  0.43$^{+0.15}_{-0.09}$ \\ 
    	\cline{1-10}
        Host M & & & & & & & & & \\
    	\cline{1-10}
        K2-25       &    & 13.9  & 0.26$\pm$0.02 & 1.14$^{+0.11}_{-0.09}$ &-  &6.28$^{+0.63}_{-0.62}$&5.47$^{+1.00}_{-0.77}$& -& 8.61$^{+0.50}_{-0.45}$\\
        GJ 436	    & A  & 3.0 & 0.11$\pm$0.02  &  0.31$^{+0.07}_{-0.05}$ & - &  0.15$^{+0.08}_{-0.04}$    &  0.051$^{+0.018}_{-0.016}$ & -    &  2.37$^{+0.15}_{-0.54}$     \\
                    & B  & 3.0  &   0.07$^{+0.04}_{-0.03}$ & 0.75$^{+0.15}_{-0.16}$ & - & 0.93$^{+0.05}_{-0.83}$& 0.018$\pm$0.005 & - & 2.29$^{+0.26}_{-1.79}$ \\
                    & C  & 3.0  & 0.19$^{+0.03}_{-0.04}$ & 2.12$^{+0.70}_{-2.12}$  & -& 0.10$^{+0.06}_{-0.03}$ &0.050$^{+0.081}_{-0.022}$ & - &  1.80$^{+0.33}_{-0.45}$    \\
                    & D  & 3.0  & 0.23$\pm$0.5 & 1.21$^{+0.17}_{-0.14}$ & - & 0.059$^{+0.027}_{-0.021}$& 0.054$^{+0.016}_{-0.013}$&-  &   1.60 $^{+0.22}_{-0.36}$ \\ 
                    & E  & 3.0  & 0.24$\pm$0.05 & 0.93$^{+0.26}_{-0.19}$ & - &  0.10$^{+0.04}_{-0.03}$ & 0.025$^{+0.015}_{-0.014}$ & - &   1.97$^{+0.37}_{-0.43}$  \\            
                    & F  & 3.0  & 0.35$^{+0.39}_{-0.07}$  & - & - & 0.11$\pm$0.06 & -& -& 1.79$^{+0.34}_{-0.63}$  \\
                    & G  & 3.0  & 0.27$^{+0.04}_{-0.03}$ & - &- & 0.20$^{+0.05}_{-0.04}$& -& - &   2.86$^{+0.53}_{-0.50}$   \\
                    & H  & 3.0& 0.12$\pm$0.02 & 0.53$^{+0.08}_{-0.09}$ & - & 0.15$^{+0.04}_{-0.03}$    & 0.055$^{+0.012}_{-0.008}$  &- &   2.93$^{+0.20}_{-0.33}$   \\
        LHS 1140    &    & 4.6   & 0.15$\pm$0.03 & 1.14$^{+0.50}_{-0.24}$ &-  & 0.068$^{+0.031}_{-0.016}$   &    0.18$^{+0.02}_{-0.01}$ & -  & 0.50$^{+0.07}_{-0.08}$     \\  
        AU Mic      & A  & 3.0   &    2.60$^{+0.09}_{-0.07}$    &  0.916$^{+0.003}_{-0.004}$   &    0.240$\pm$0.001         &  38.01$^{+0.59}_{-0.37}$  & 38.37$^{+0.50}_{-0.34}$ & 71.69$^{+0.22}_{-0.35}$ & 2209$\pm$4\\
                    & B  & 3.0 &3.17$^{+0.05}_{-0.04}$ & 0.963$^{+0.002}_{-0.001}$  & 0.238$^{+0.005}_{-0.004}$ & 45.43$^{+0.22}_{-0.23}$    & 77.27$^{+0.34}_{-0.45}$ & 76.24$\pm$0.34 & 2717$^{+5}_{-2}$  \\
                    & C  & 3.0  &   3.19$^{+0.04}_{-0.06}$        &   0.966$^{+0.001}_{-0.002}$          & 0.239$\pm$0.001   & 67.25$^{+0.50}_{-0.33}$  & 44.99$^{+0.16}_{-0.24}$ & 76.62$^{+0.32}_{-0.20}$ & 2617$^{+2}_{-3}$   \\
                    & D  &  3.0 &   3.17$\pm$0.06        &  0.953$\pm$0.002           & 0.236$\pm$0.001    &   54.92$\pm$0.34 & 39.83$\pm$0.23 & 72.00$\pm$0.23& 2351$^{+1}_{-3}$ \\
                    & E  & 3.0  & 3.18$^{+0.06}_{-0.07}$           &  0.974$\pm$0.003            & 0.237$\pm$0.001   &  97.58$\pm$0.68 &  54.36$^{+0.34}_{-0.45}$ & 83.99$\pm$0.34 & 3164$^{+4}_{-5}$  \\  
        GJ 3470     &    &  9.1   & 0.28$\pm$0.02  &  1.45$^{+0.60}_{-0.30}$   & -  &    1.71$^{+0.16}_{-0.15}$      &    0.62$^{+0.26}_{-0.19}$   & -   & 3.68$^{+0.31}_{-0.25}$                 \\
        GJ 1214     & A  & 4.5 & 1.71$^{+0.86}_{-0.35}$  & 0.25$^{+0.06}_{-0.05}$ & -& 0.25$^{+0.16}_{-0.11}$   & 0.20$\pm$0.07    & - & 0.62$^{+0.27}_{-0.06}$  \\
                    & B  & 4.5 & 0.30 & -& -& 0.05& -& -& $<0.32$\\[2pt]
    	\hline
\end{tabular}
\end{table*}

\section{X-ray luminosities}
\label{sec:x}
The targets were selected for having publicly available X-ray observations with either XMM-Newton or Chandra (or both). 
Table \ref{tab:x2} and Table \ref{tab:x3} (the latter is specific to HD~189733, which was observed 31 times) list the full observations' log. 

XMM-EPIC-pn (European Photon Imaging Camera) data were analyzed using the Scientific Analysis System (SAS 15.0.0), while Chandra ACIS-S (Advanced CCD Imaging Spectrometer) and HRC (High Resolution Camera) observations were reduced with the Chandra Interactive Analysis of Observations software package (CIAO 4.13). For each observation, we estimated the source sky position using J2000 coordinates and proper motion provided by SIMBAD\footnote{\url{http://simbad.u-strasbg.fr/simbad/}}. An X-ray source was detected within 1.0 arcsec of the target star optical position (after accounting for proper motion) with a significance greater than $3\sigma$ \citep{Li1983} in at least 1 observation for 22 out of 26 target stars. 

Photometry was performed within a radius of $\sim$15 and $\sim$2 arcsec for EPIC and ACIS/HRC, respectively, and centered on the nominal optical position of the target star, corrected for proper motion. For the XMM observations of HD 189733 and LHS 1140 we adopted a 10 arcsec extraction radius, to avoid contamination from nearby sources \citep{Pillitteri2010,Spinelli2019}. The background was extracted from a contamination-free circular region with a radius of about 30 arcsec for all targets. 

We analysed the background-corrected spectra with XSPEC 12.11.1 \citep{Arnaud1996}. Owing to the low numbers of counts (the median of the net counts of the sources is 106 counts), each spectrum was binned to have at least one count per energy bin. The resulting spectra were fitted using C-statistics (hereafter cstat; \citealt{Cash1979}), and assuming a thermal plasma emission model. 
Metal abundances were fixed to the solar value as derived by \citet{Asplund2009} (in no case did leaving the abundance free to vary improve the fit). 

We modeled the X-ray flux attenuation arising from interstellar absorption using the TBABS model \citep{Wilms2000}. 
For each system, the hydrogen column density was set equal to the product between the distance and a mean interstellar hydrogen density of 0.1 cm$^{-3}$, in agreement with LIC model (\citealt{Linsky2000, Redfield2008}\footnote{\url{http://lism.wesleyan.edu/ColoradoLIC.html}}). 

\begin{figure*} 
\centering
\includegraphics[width=0.24\textwidth]{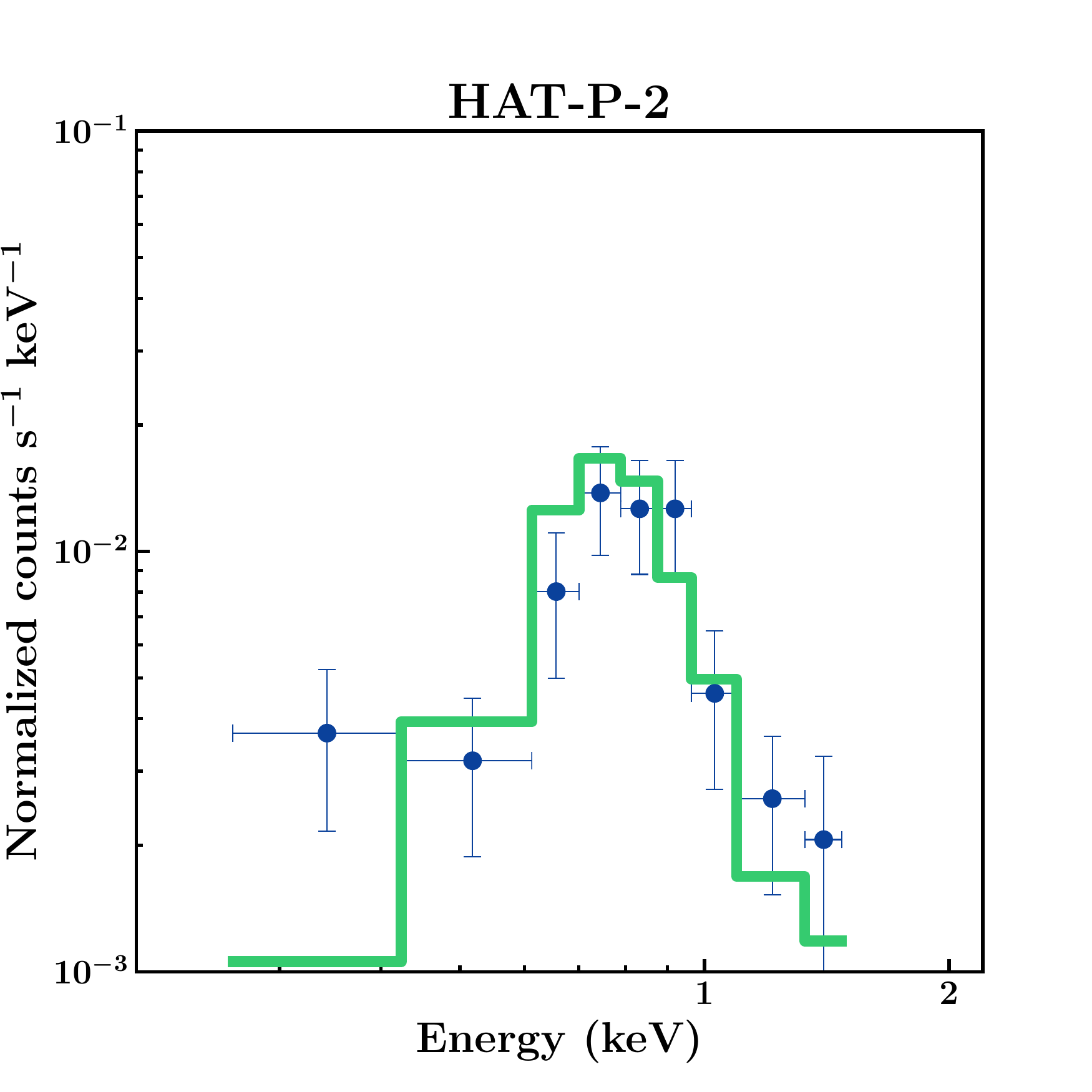} 
\includegraphics[width=0.24\textwidth]{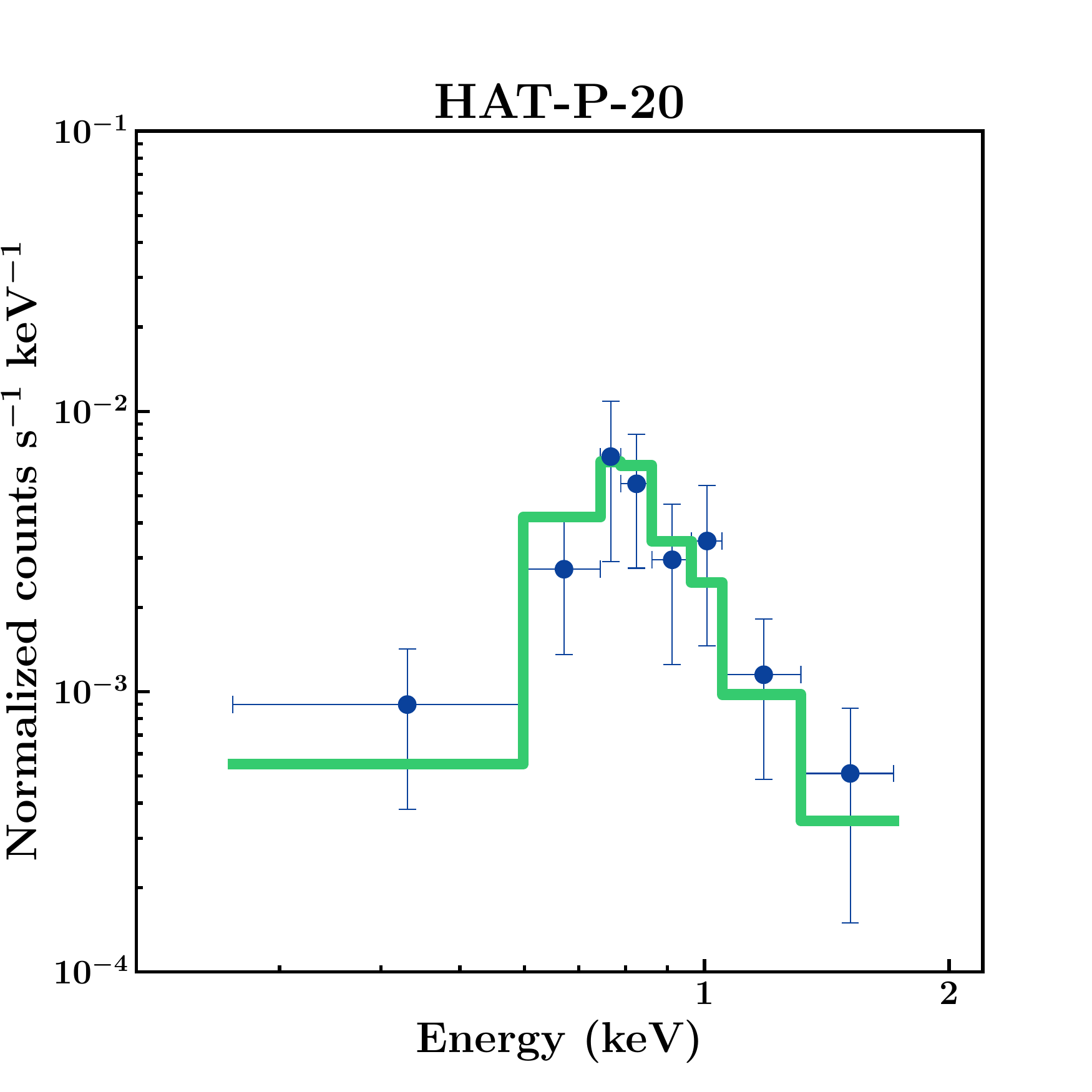}
\includegraphics[width=0.24\textwidth]{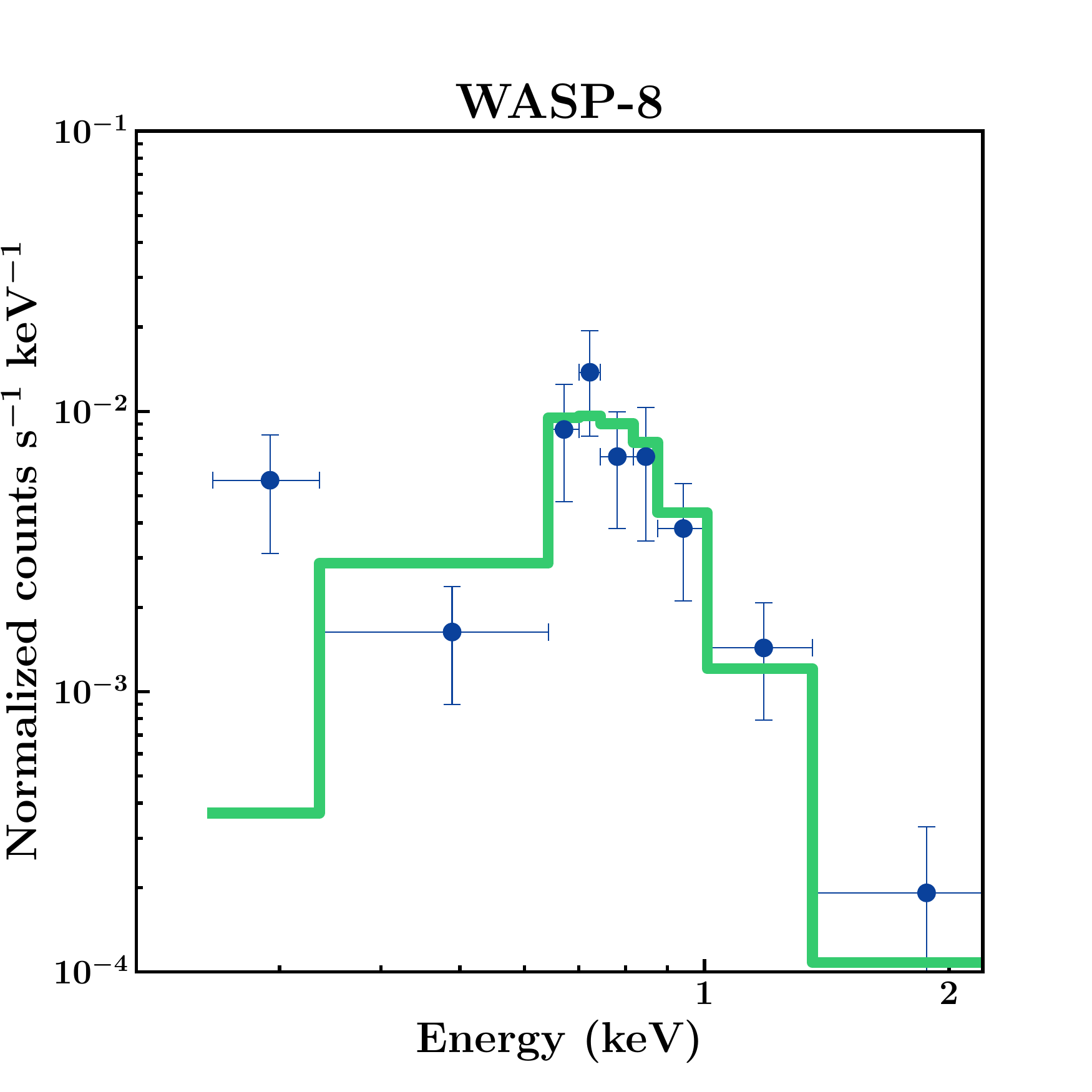}
\includegraphics[width=0.24\textwidth]{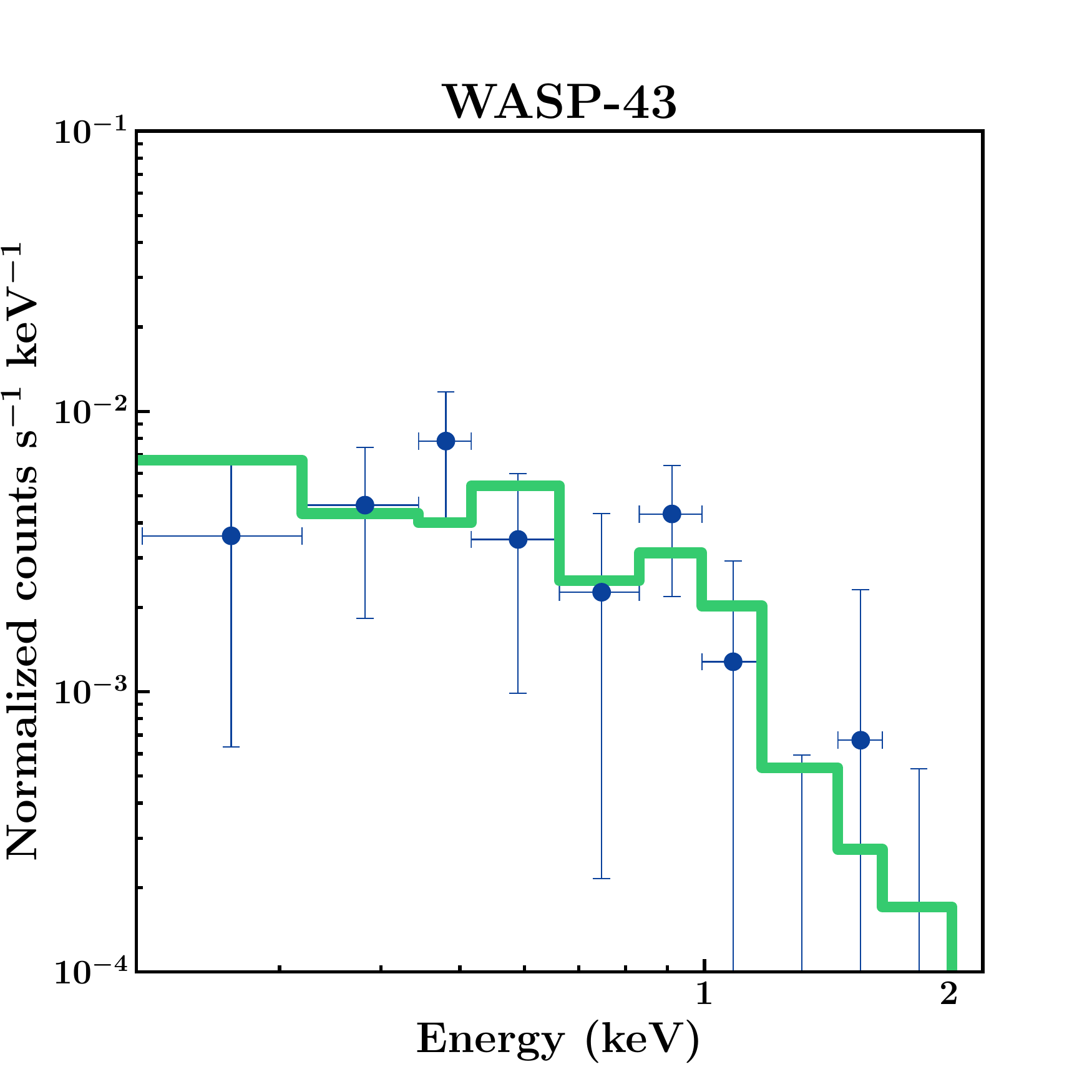}\\
\includegraphics[width=0.24\textwidth]{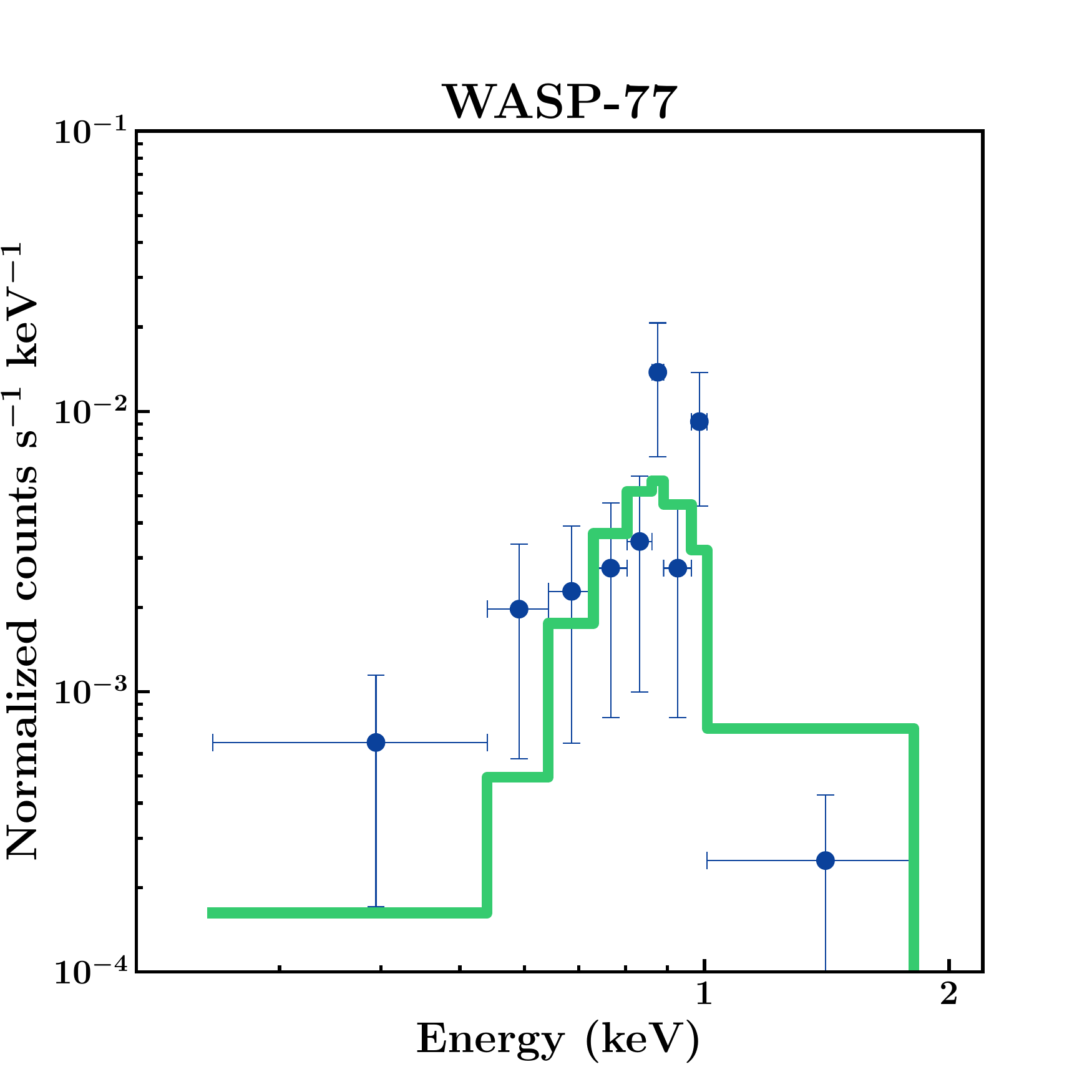}
\includegraphics[width=0.24\textwidth]{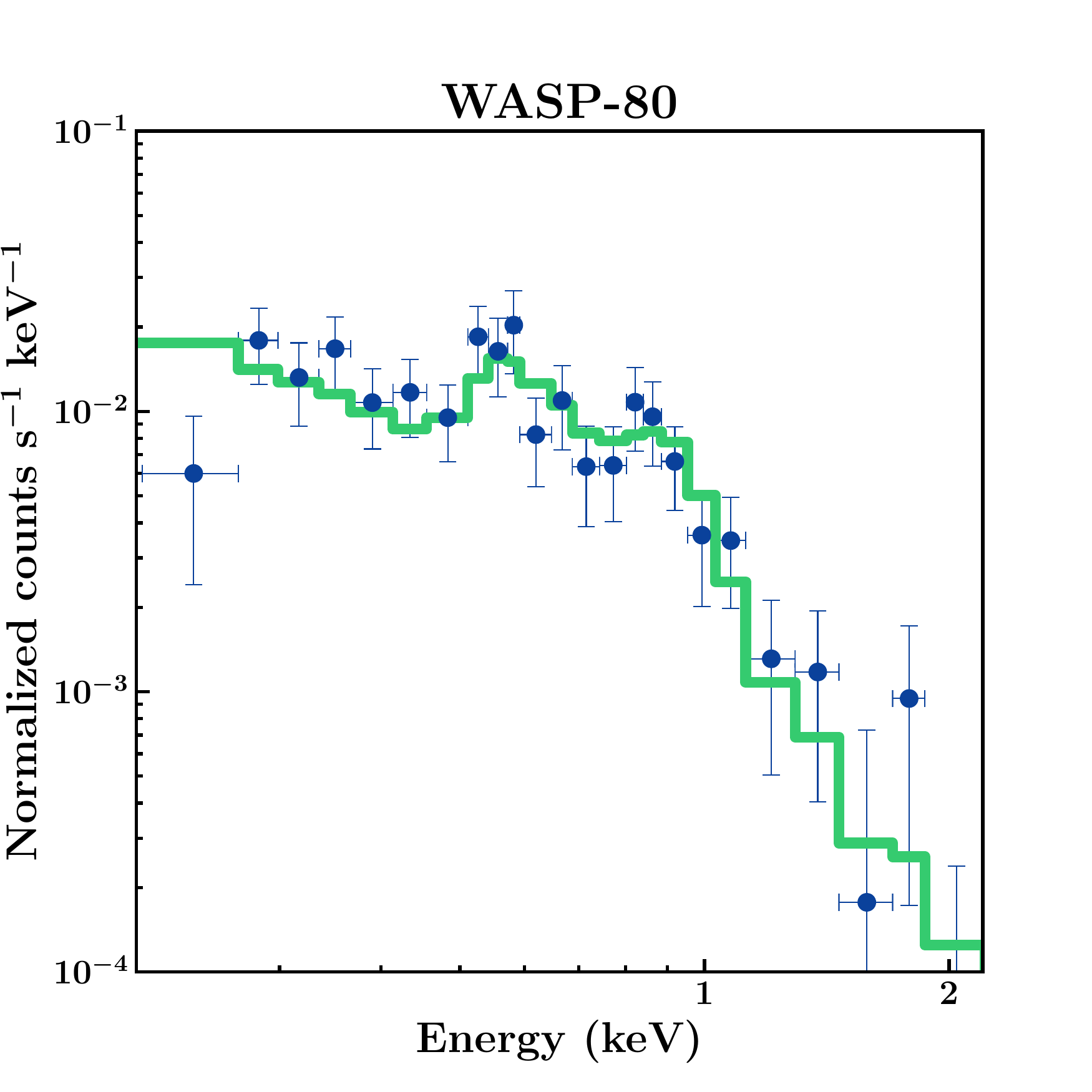} 
\includegraphics[width=0.24\textwidth]{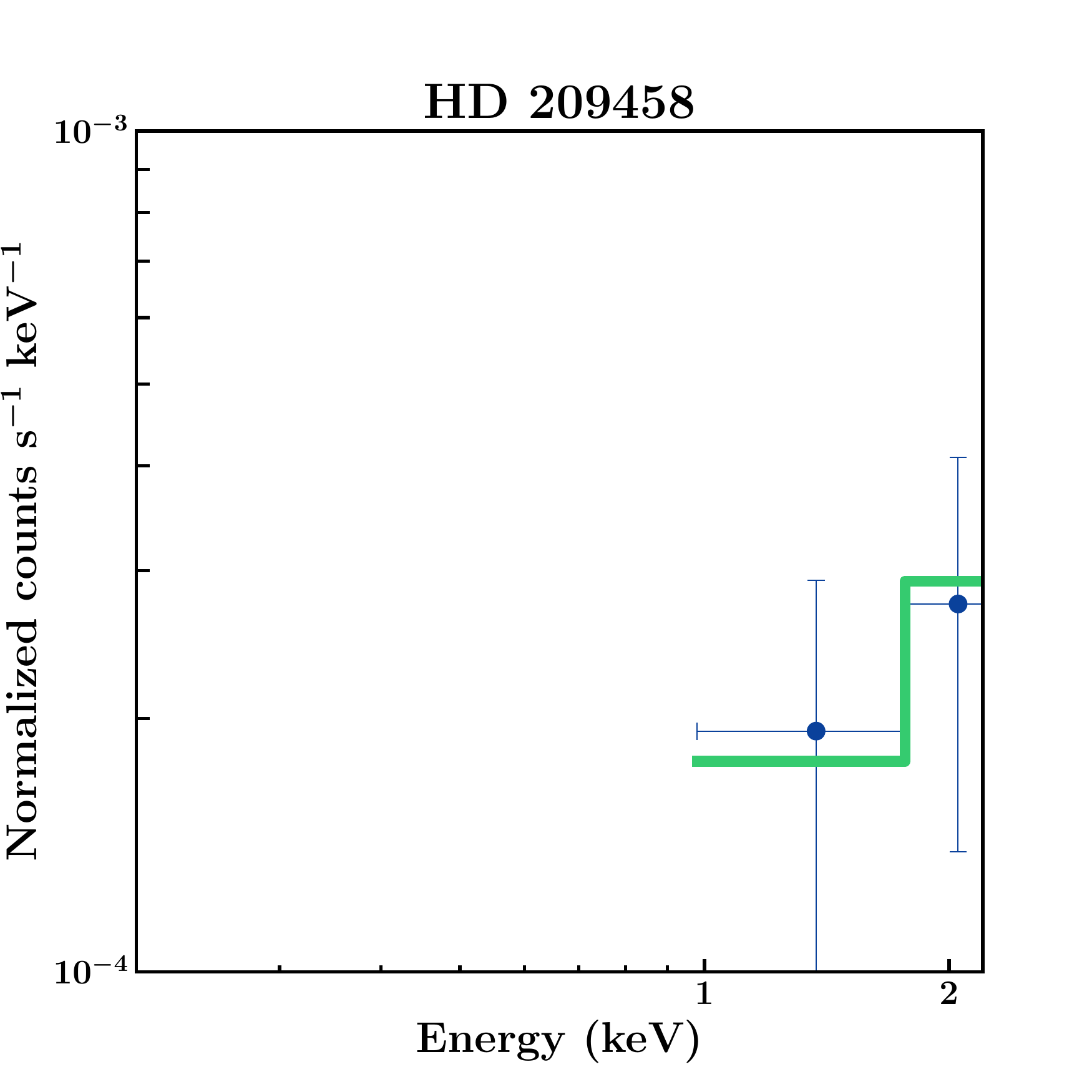}
\includegraphics[width=0.24\textwidth]{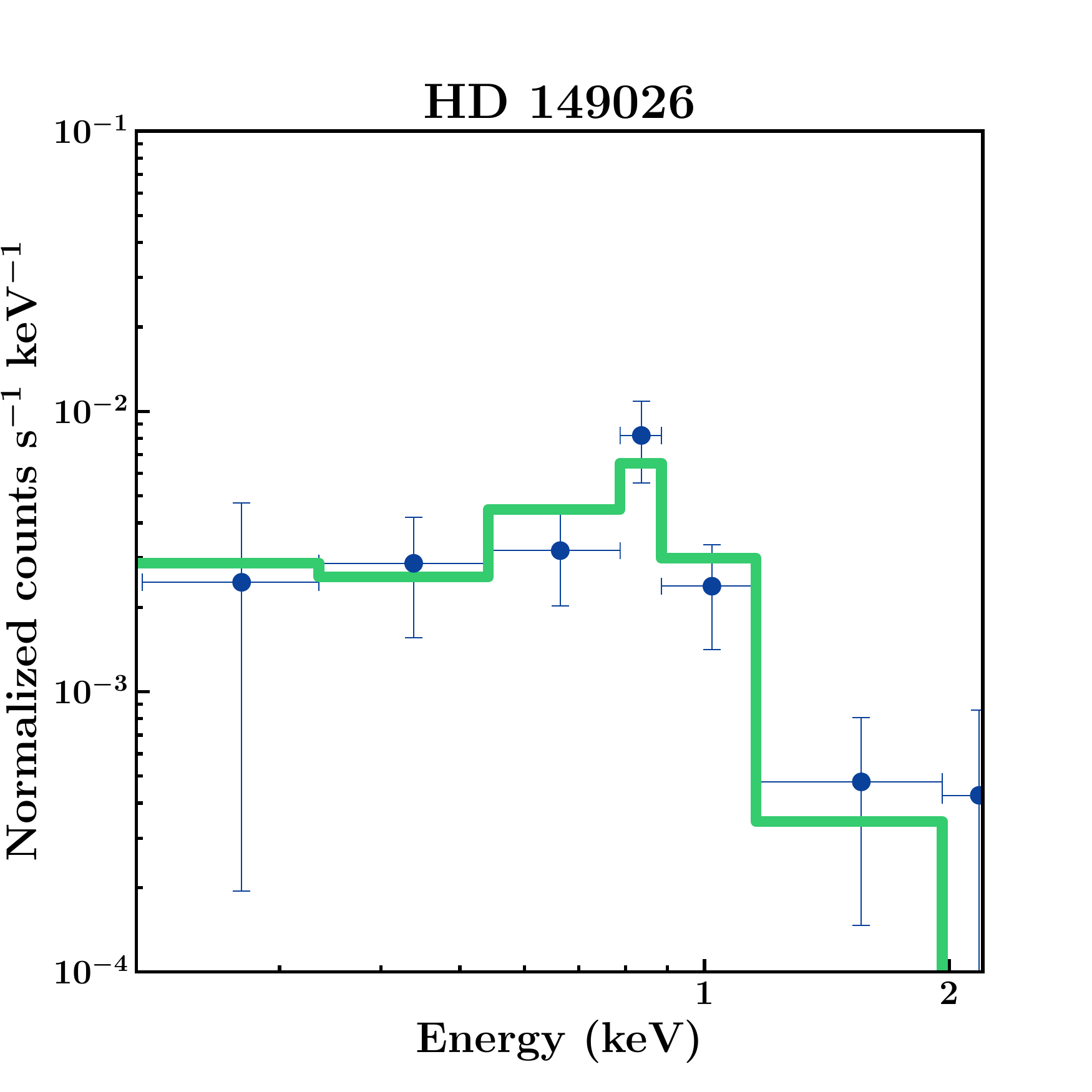}\\
\includegraphics[width=0.24\textwidth]{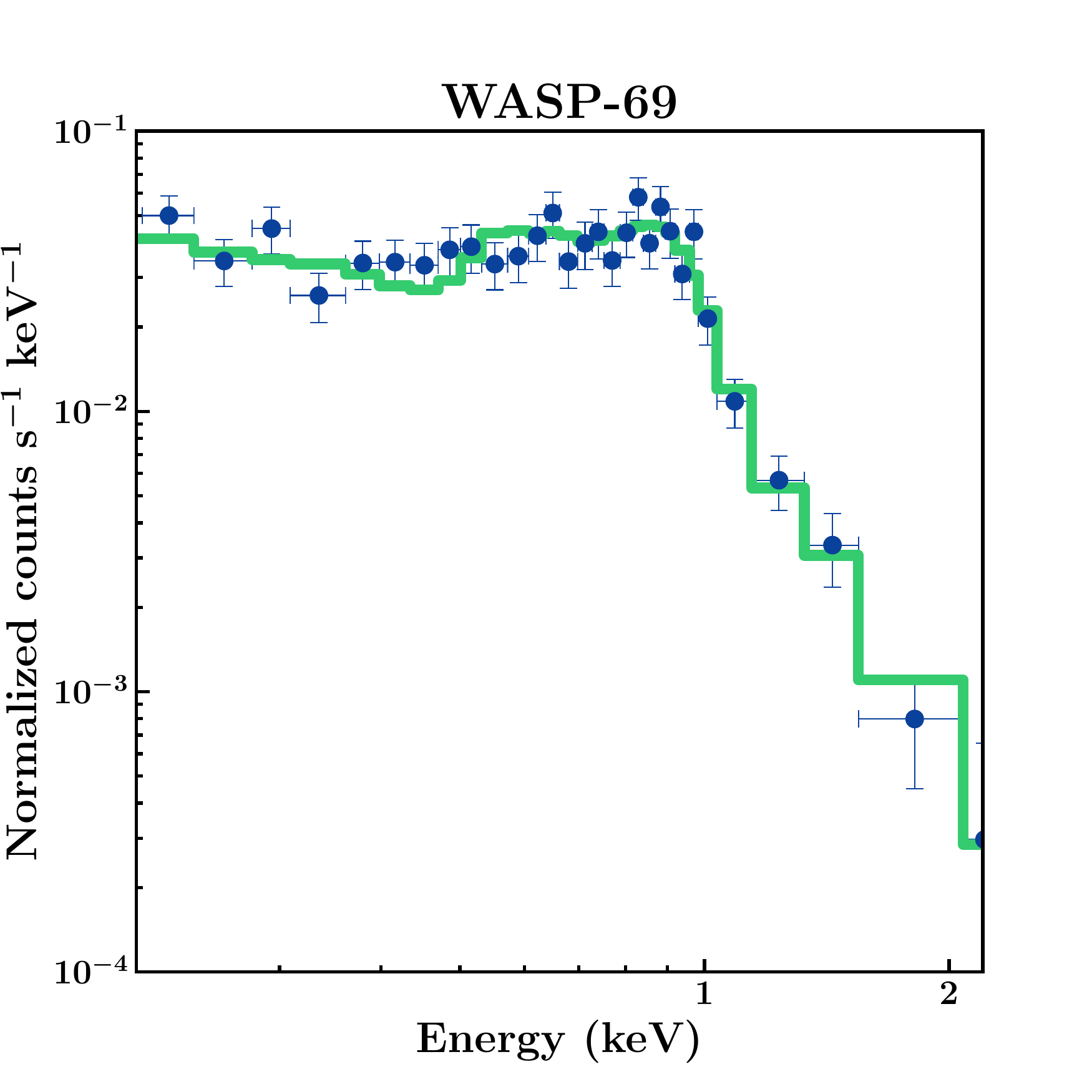}
\includegraphics[width=0.24\textwidth]{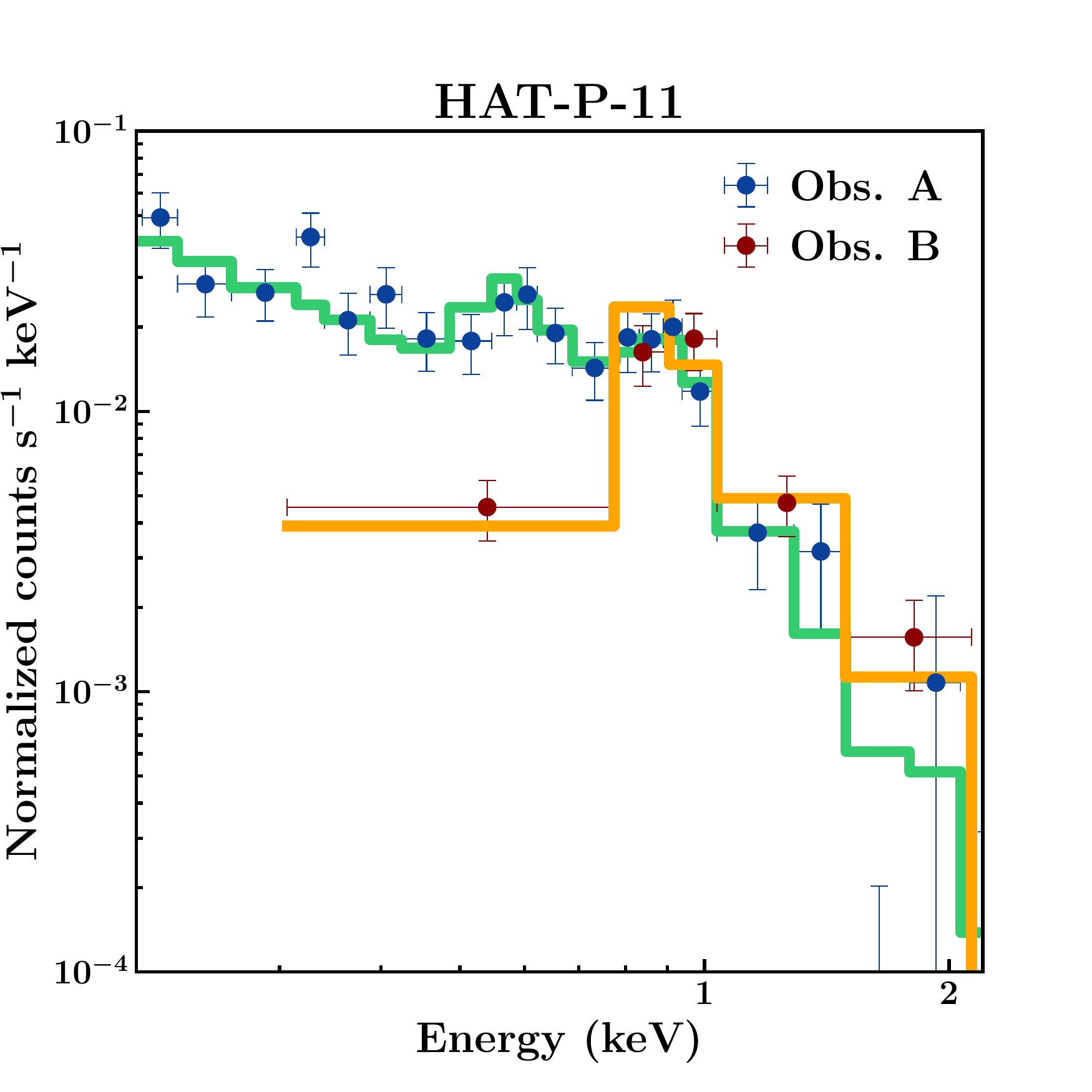} 
\includegraphics[width=0.24\textwidth]{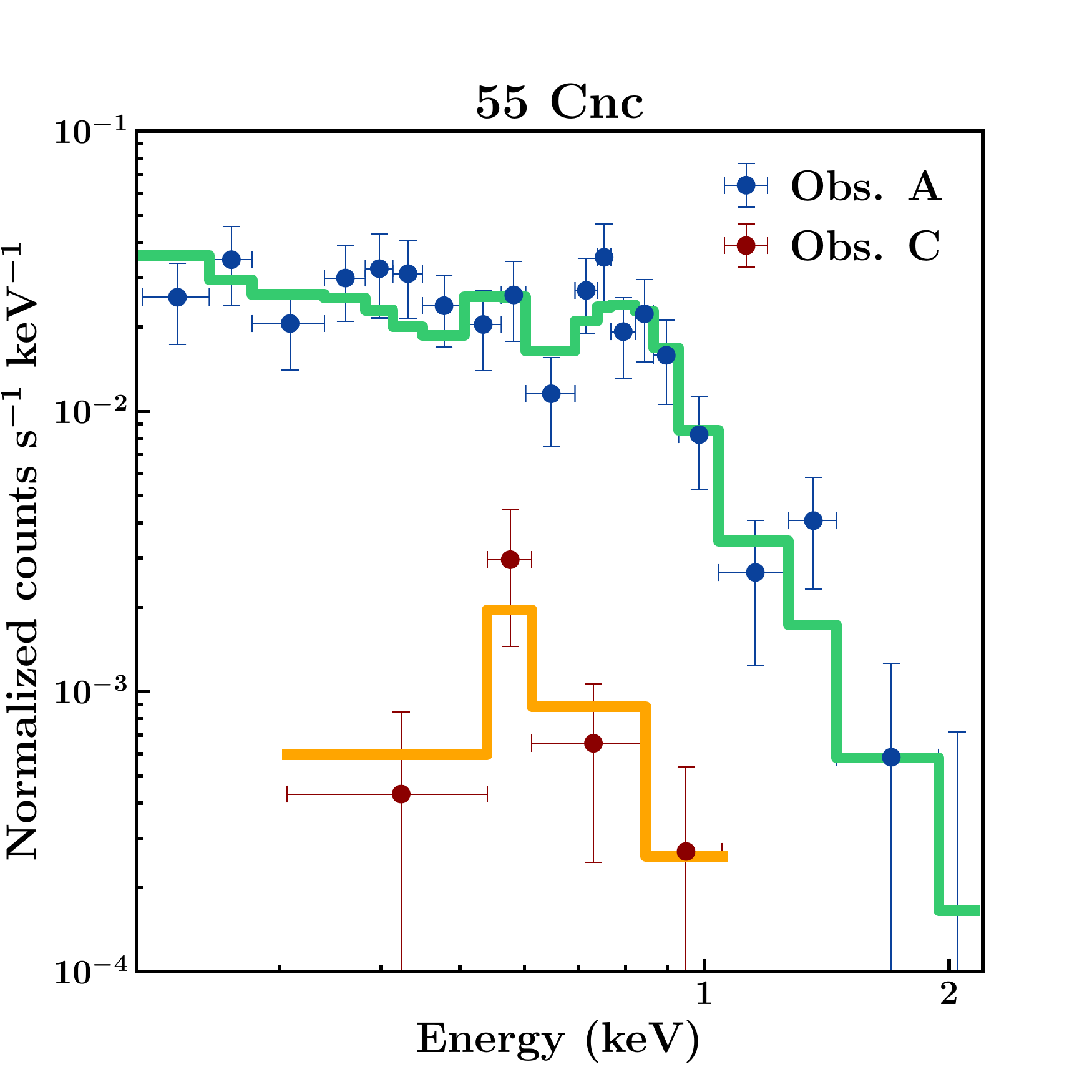} 
\includegraphics[width=0.24\textwidth]{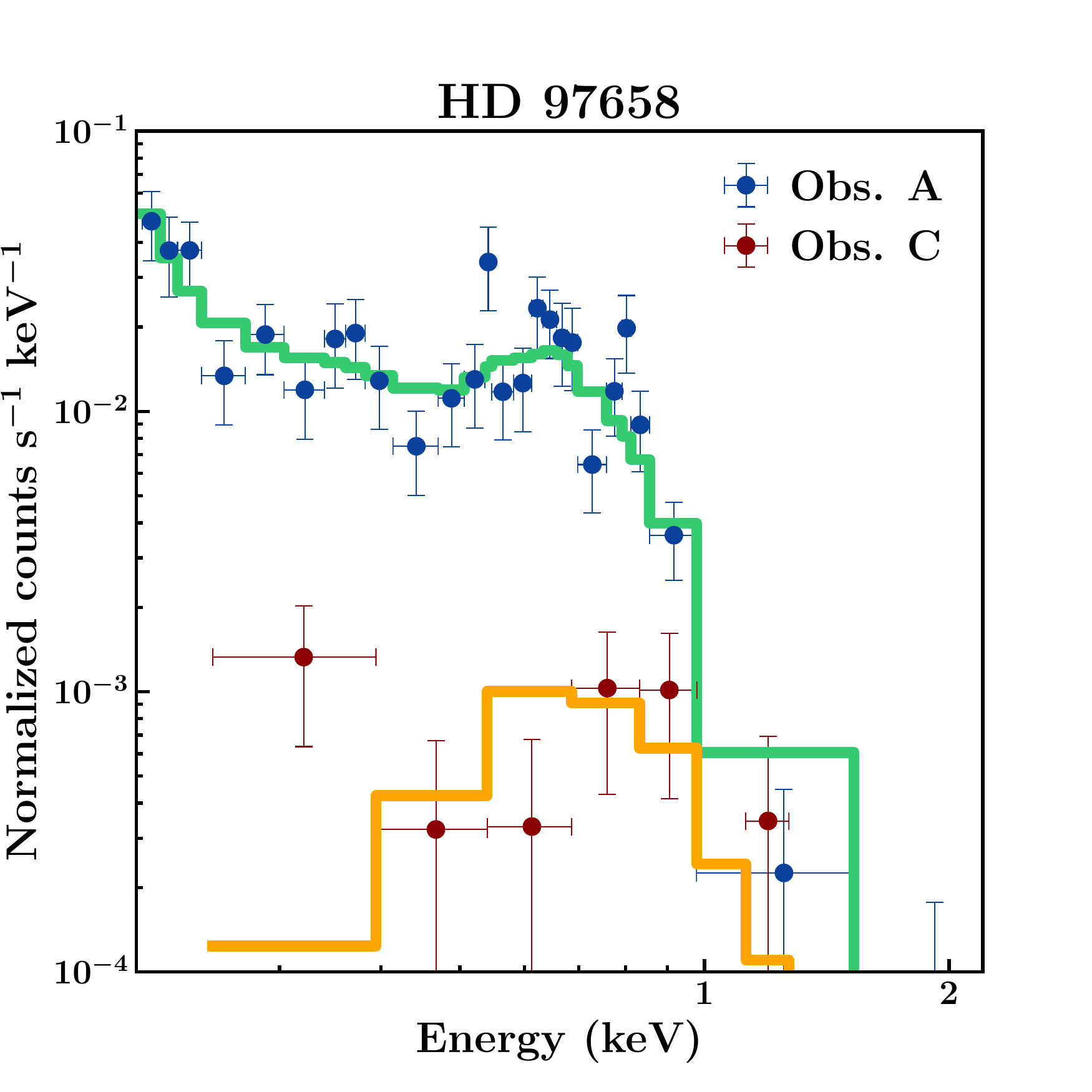} \\
\includegraphics[width=0.24\textwidth]{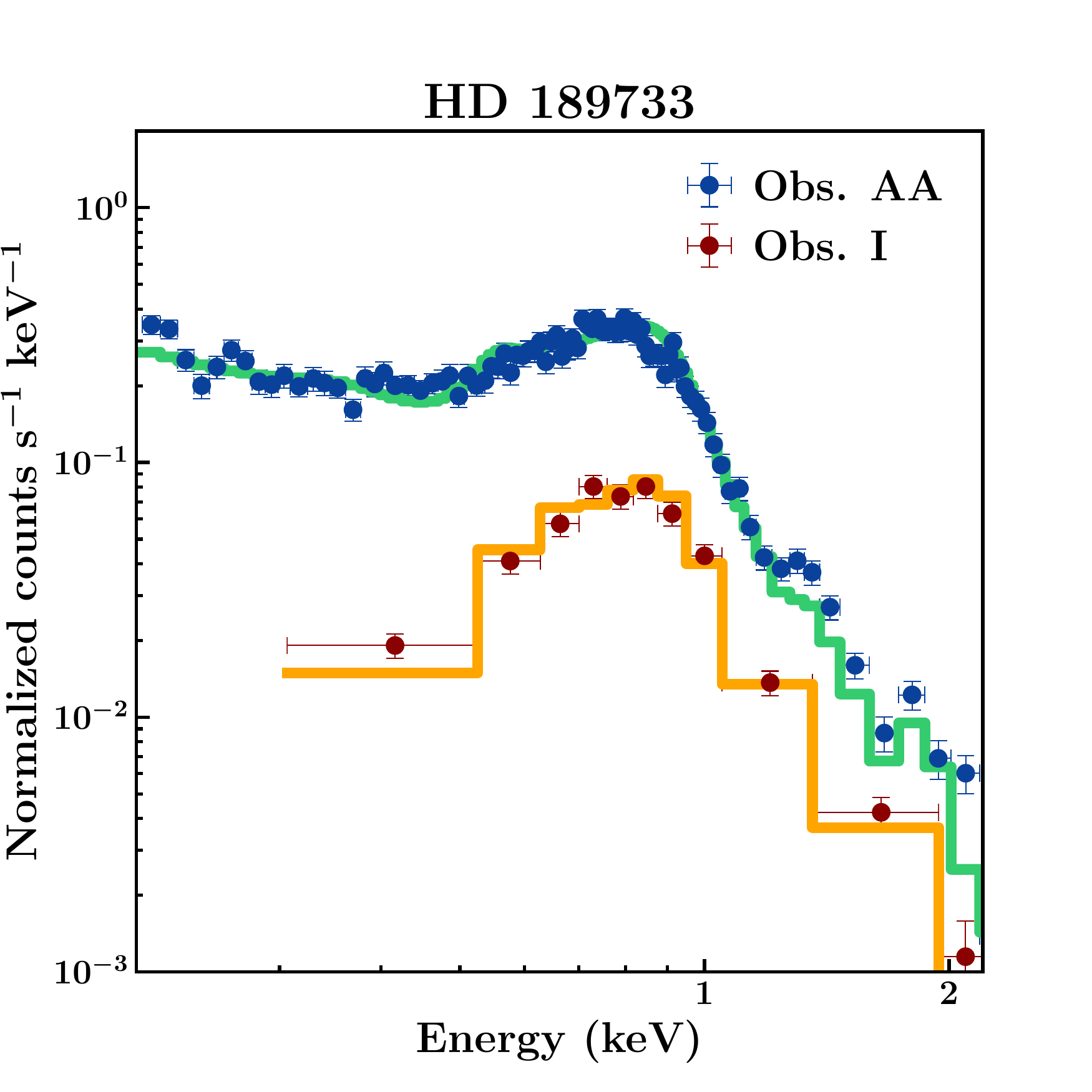} 
\includegraphics[width=0.24\textwidth]{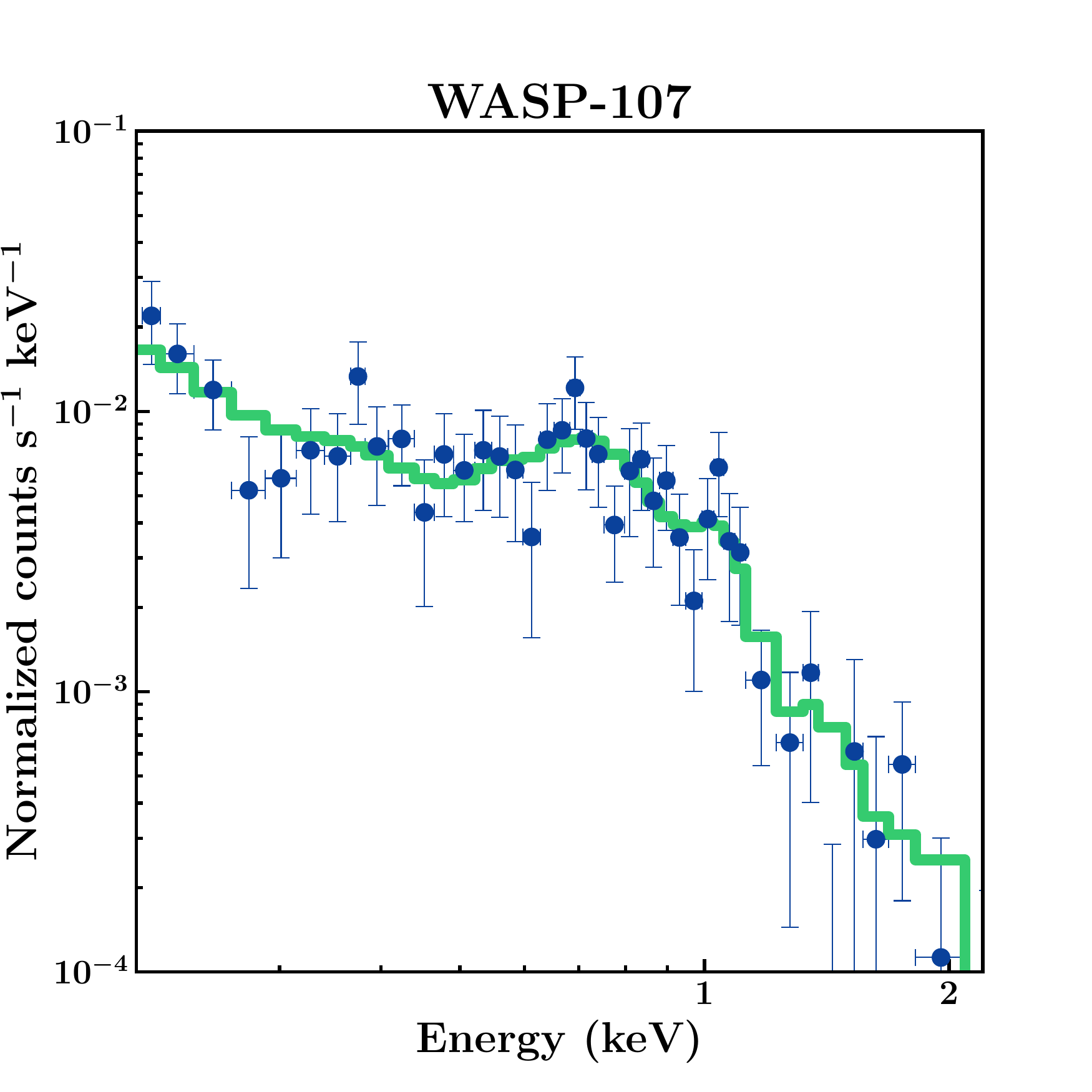} 
\includegraphics[width=0.24\textwidth]{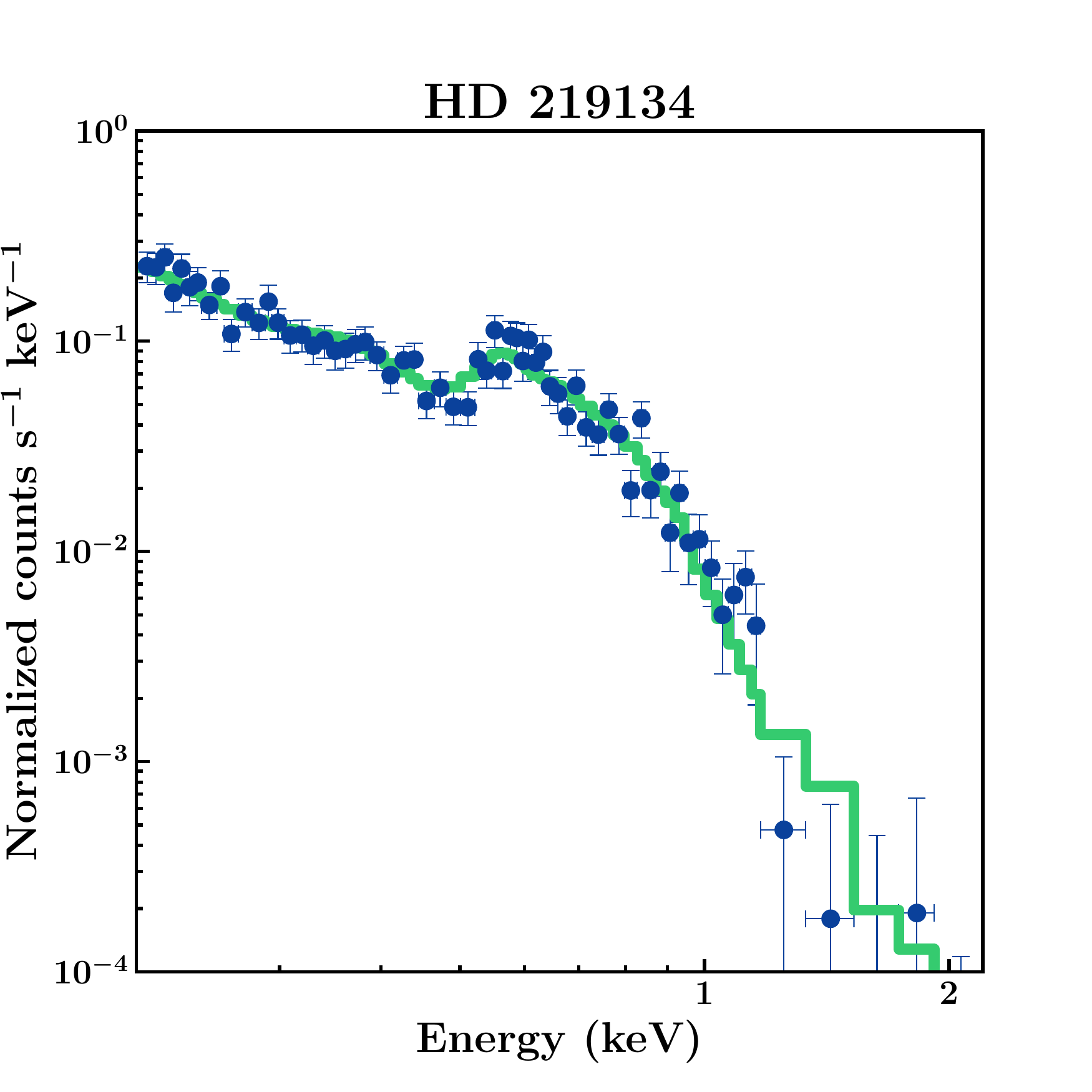} 
\includegraphics[width=0.24\textwidth]{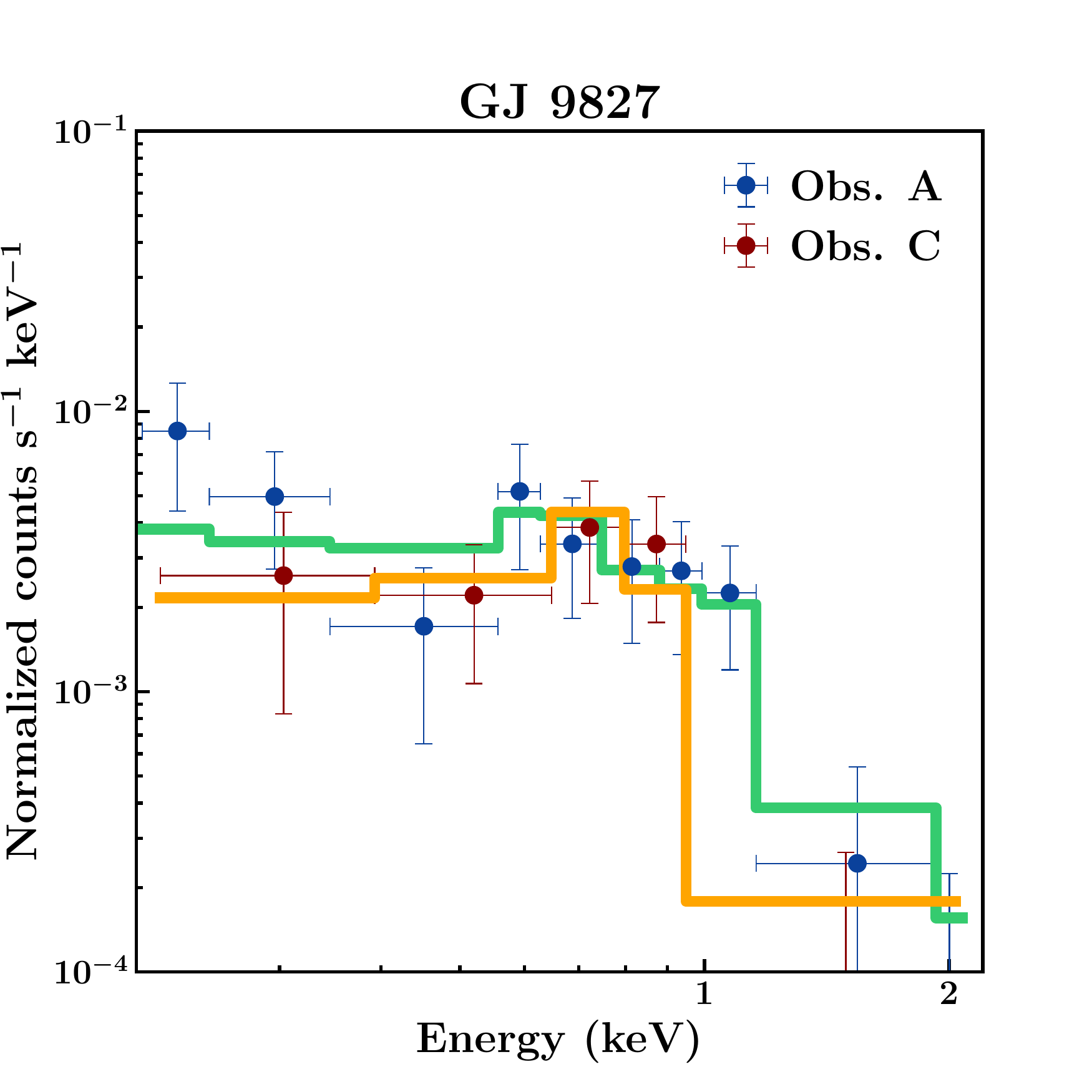} 
\caption{EPIC-pn and ACIS-S (binned) X-ray spectra for the X-ray detected FKG-type host stars. Count rates are shown in blue/red with error bars; the green/orange histograms represent the best-fit (one, two, or three temperature) APEC model. For targets with multiple epochs and significant variability we show the highest and the lower flux cases.}
\label{fig:2}
\end{figure*}

\begin{figure*} 
\centering
\includegraphics[width=0.24\textwidth]{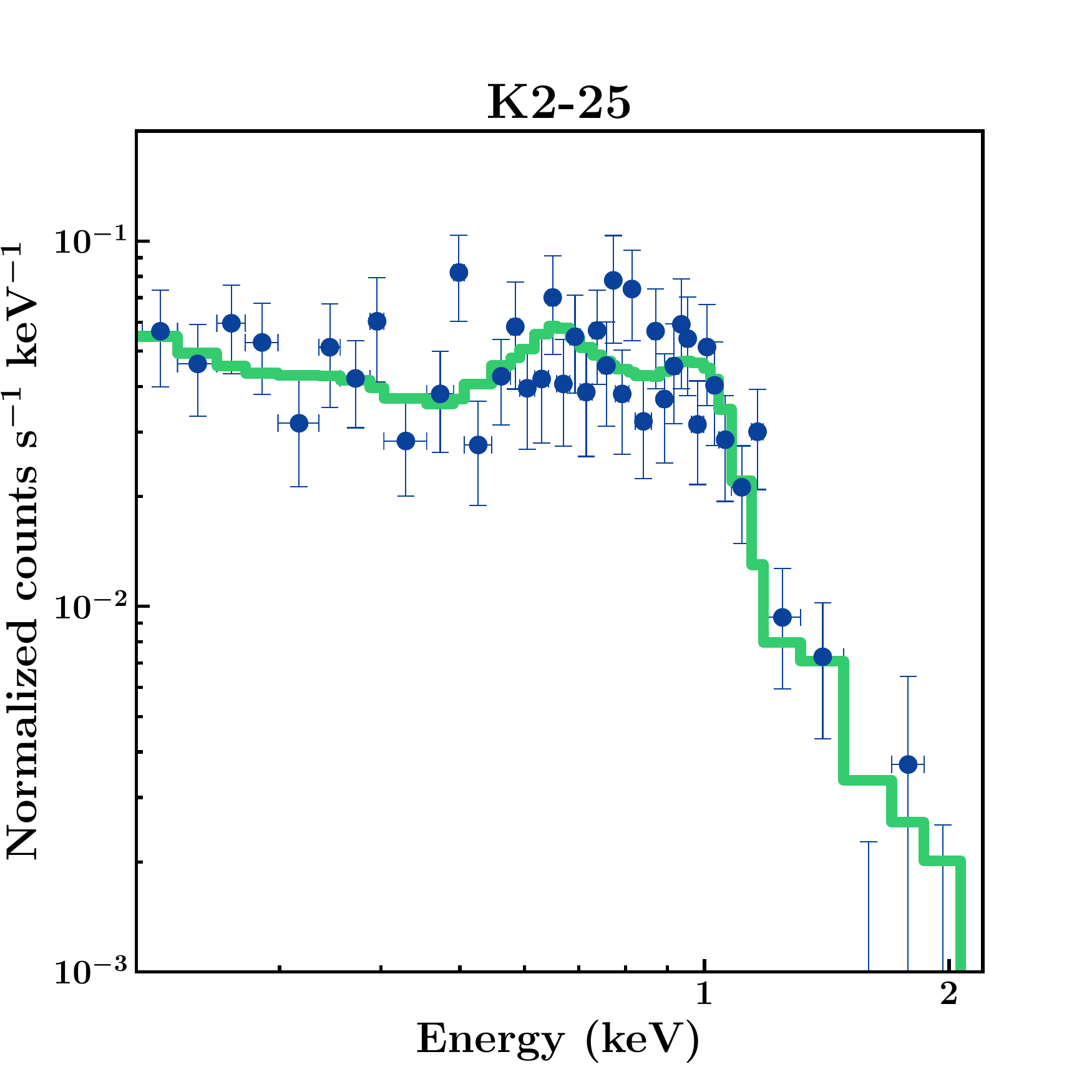}
\includegraphics[width=0.24\textwidth]{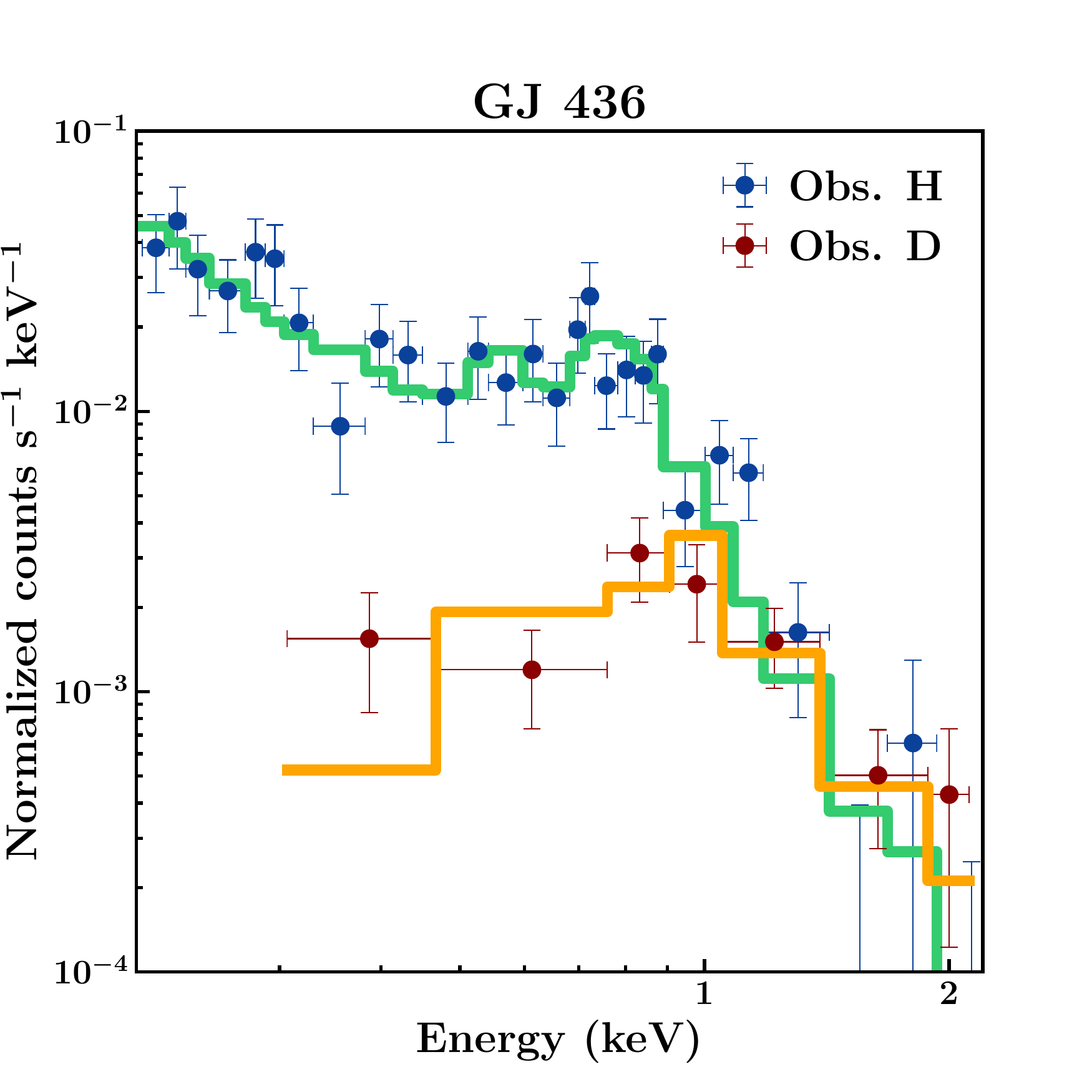} 
\includegraphics[width=0.24\textwidth]{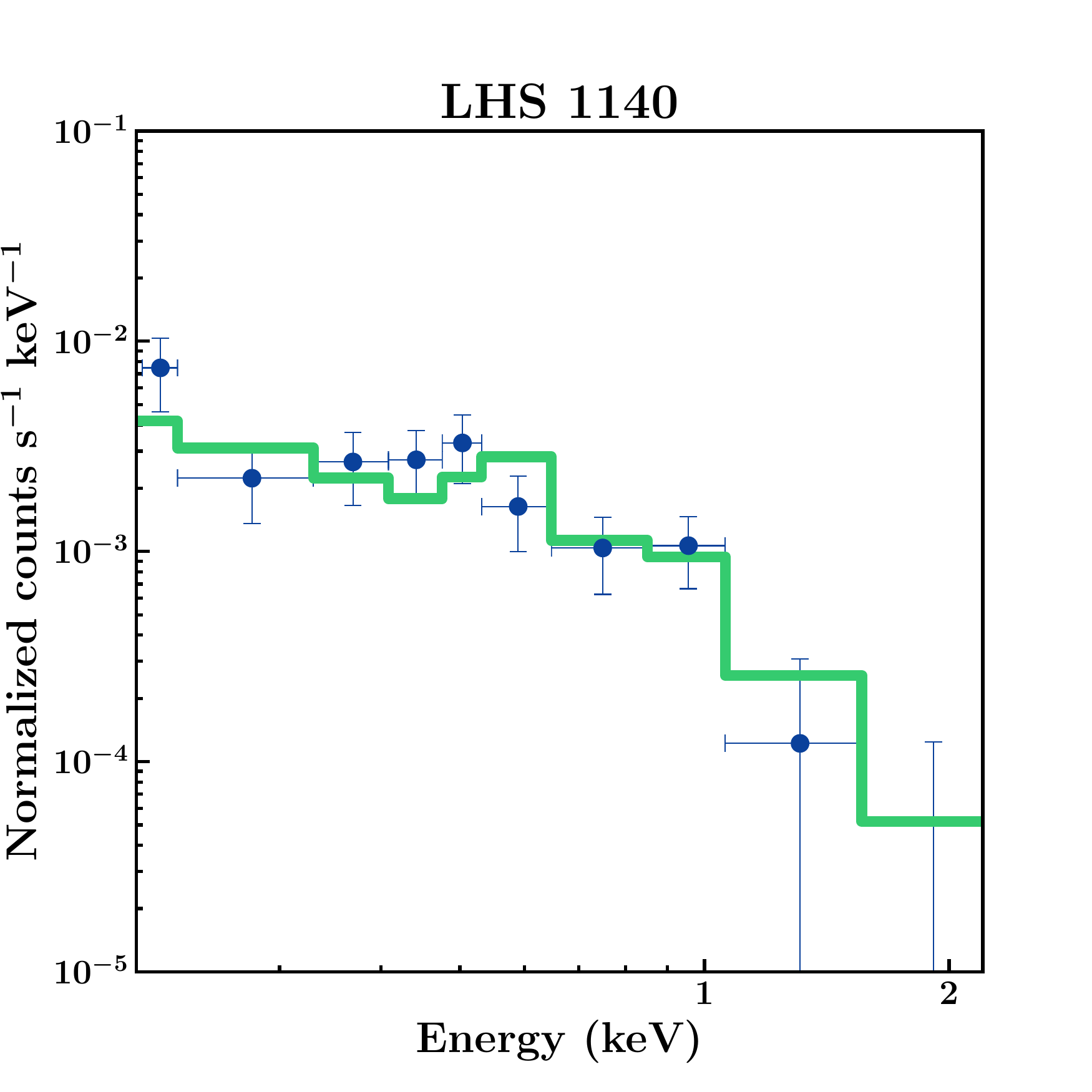} \\
\includegraphics[width=0.24\textwidth]{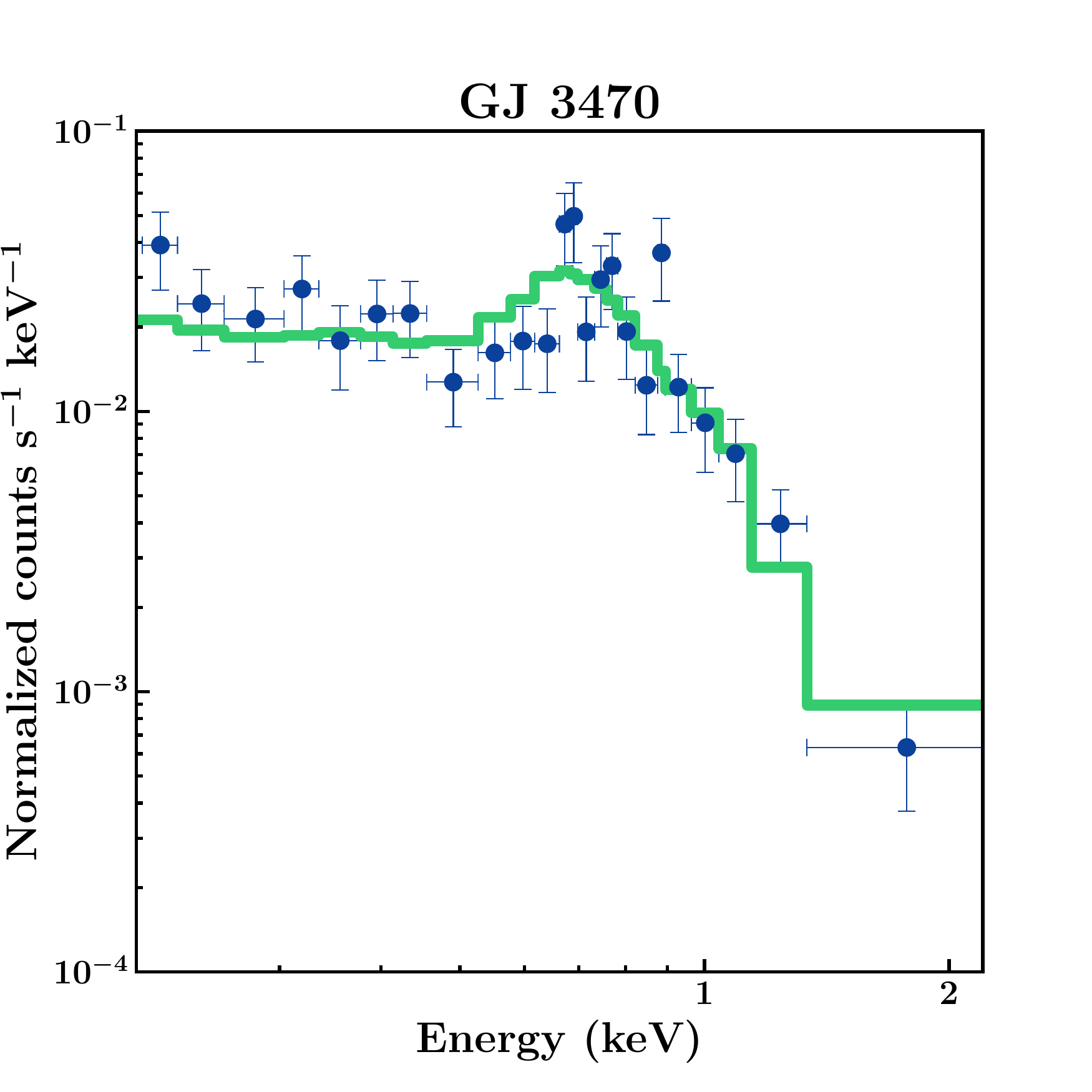} 
\includegraphics[width=0.24\textwidth]{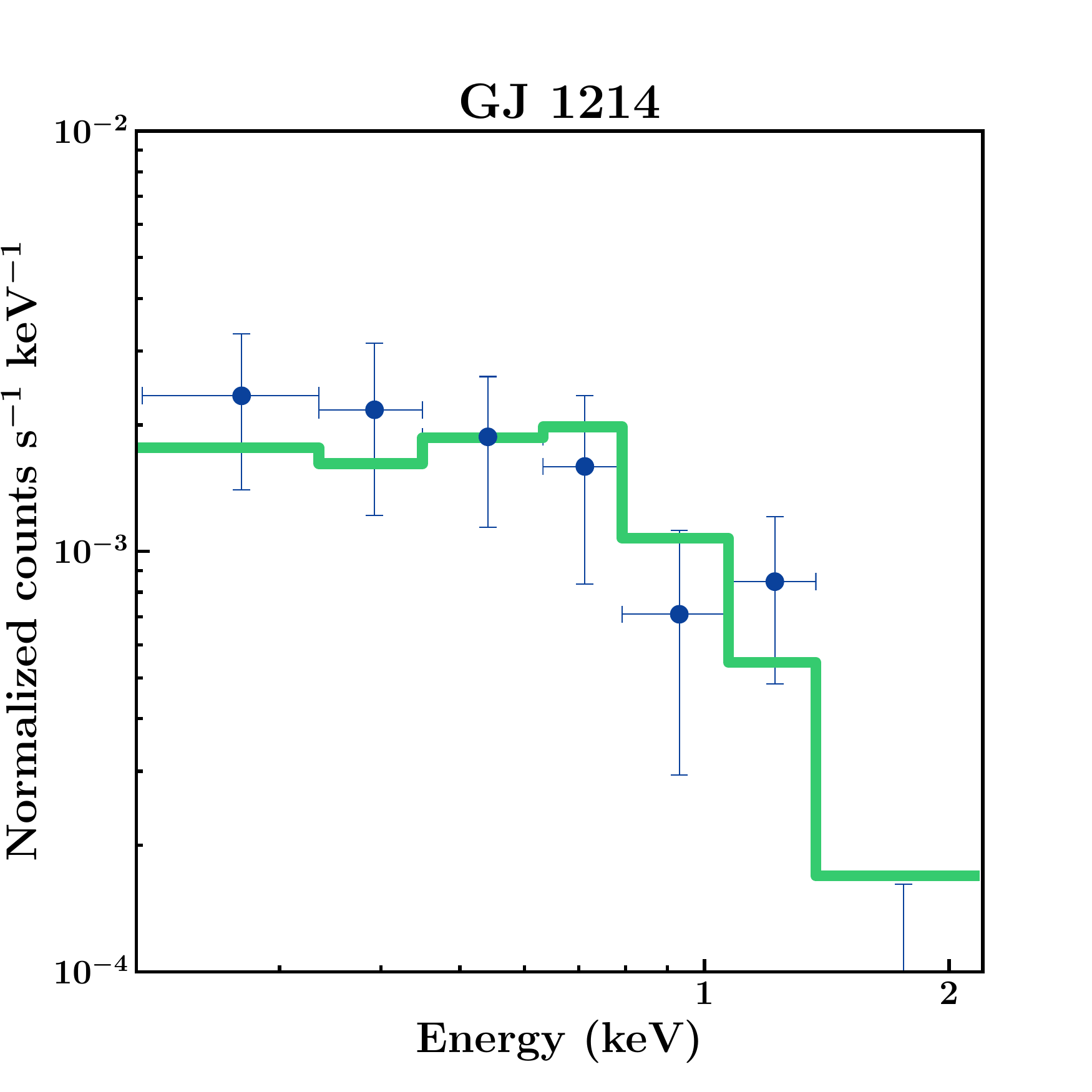} 
\includegraphics[width=0.24\textwidth]{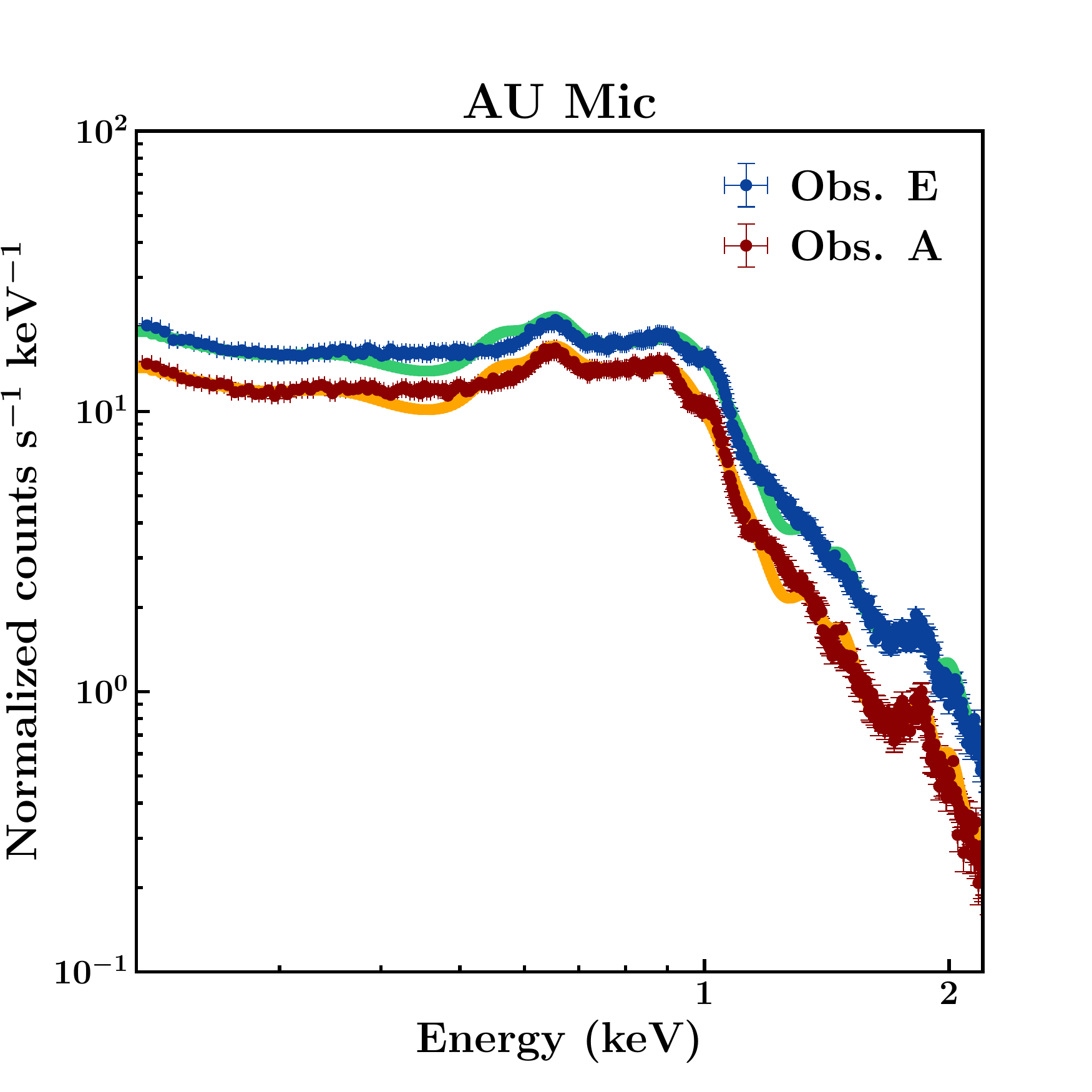} \\
\caption{Same as Figure~\ref{fig:2}, but for M-type hosts.}\label{fig:2b}
\end{figure*}

In order to assess whether a one, two or three temperature APEC model provides a better description of the data, we proceeded with a simulation, as follows. 
For each target star, we start by fitting the data with a one-temperature model, and use the resulting best-fit model to generate 1,000 spectra with the XSPEC tool \textit{fakeit}. 
If the cstat of the best-fit model to the actual data falls within 95\% the cstat distribution of the simulated data, we consider the original best-fit model as acceptable. 
We then repeat the same procedure using the two- and three-temperature APEC models. 
In those cases where more than one choice of models are acceptable, we adopt the Akaike information criterion \citep{Akaike1974}. The preferred model is identified by the AIC value $2K + C$, where $C$ is the
cstat and $K$ the number of free parameter in the model. If the difference between two AIC values is larger than 4 we choose the model with the lowest AIC; otherwise we choose the model with the lowest number of temperatures. 
All of the X-ray spectra are shown in Figures \ref{fig:2} and \ref{fig:2b} along with their best-fit models. 

For marginally significant detections (less than $3 \sigma$, i.e., HD97658-D, WASP-10, HD209458-A/B/C, WASP-38, WASP-18, WASP-29, 55Cnc-B), we estimated the $3\sigma$ confidence upper limits following \citet{Gehrels1986}. Count rate limits were converted to flux limits using  PIMMS\footnote{\url{https://cxc.harvard.edu/toolkit/pimms.jsp}} v.4.11a. 
The resulting fluxes and/or upper limits (in the energy bands 0.2-2.4 keV for XMM observations and 0.243-2.4 keV for Chandra) are listed in Table \ref{tab:flux}.\\

\begin{figure}
    \center
	\includegraphics[width=1
	\columnwidth]{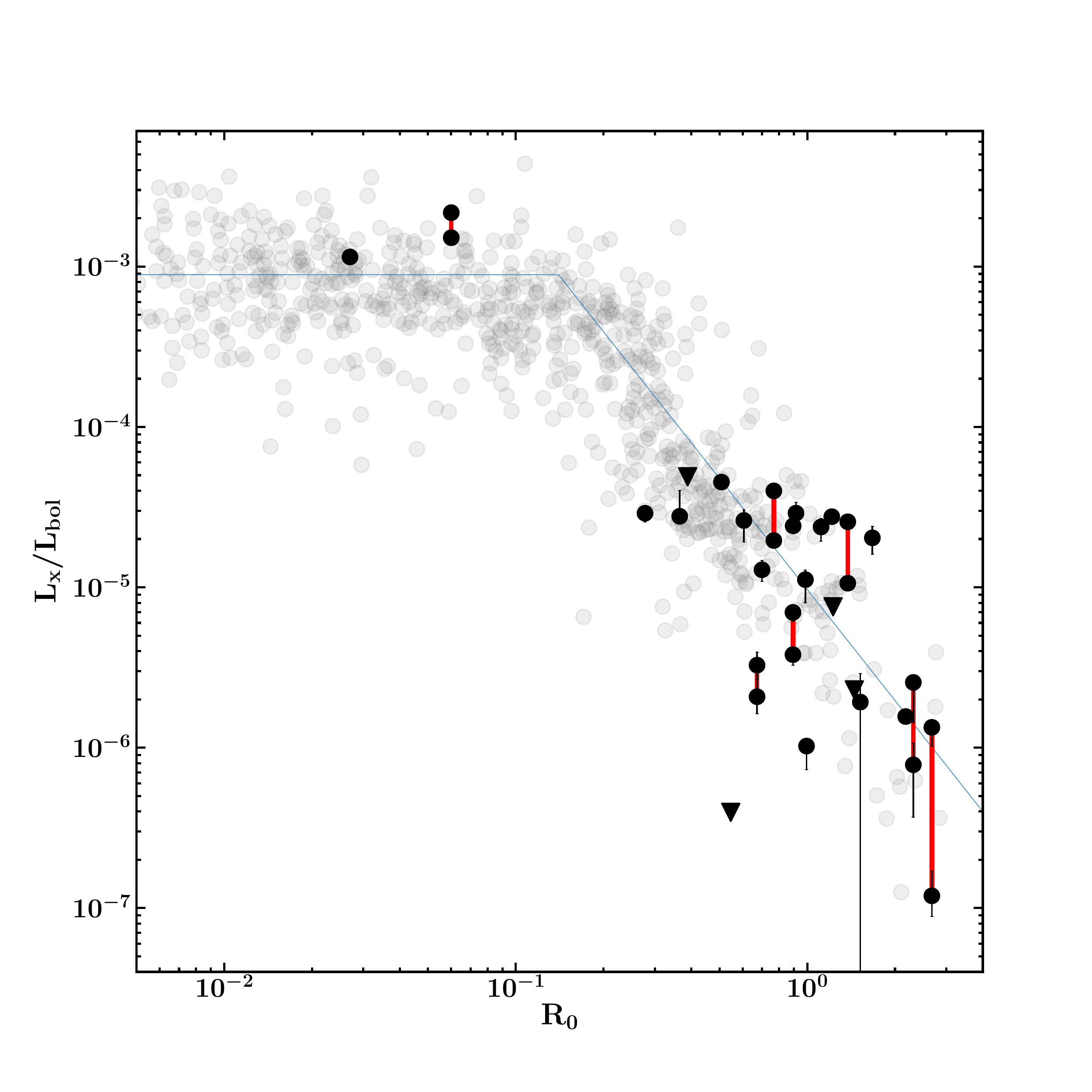}
	\caption{Stellar X-ray to bolometric luminosity ratio as a function of Rossby number, adapted from \citet{Wright2018}. The full sample of \citet{Wright2018} is shown in grey, with their best-fitting relation in blue, while the black dots/triangles refer to our target sample (triangles indicate upper limits). The red segments connect different measurements in the case of X-ray variable stars. }
	\label{fig:lxlbol}
\end{figure}

In Figure~\ref{fig:lxlbol} we overlay our X-ray measurements\footnote{Where we applied the K18 relations to convert the measured X-ray luminosities into the ROSAT energy band; adopted the bolometric luminosities estimated by \citet{Andrae2018} for FGK stars, and; used the relation of \citet{Morrell2019} to estimate $L_{\rm bol}$ for M dwarfs (using GDR2 distances). To calculate the convective turnover time which enters into the expression for the Rossby number, we used the polynomial relationship given by \cite{Wright2018}.} to the known relation between the stellar X-ray to bolometric luminosity ratio ($L_{X}/L_{\rm bol}$) and the Rossby number ($R_{\rm 0}$) (\citealt{Wright2018}, after \citealt{Wright2016} and \citealt{Wright2011}). This study, which includes fully convective stars, confirmed that very fast rotators ($R_0 < 0.13$) reach a saturation level of $L_{X}/L_{\rm bol} \simeq 10^{-3}$, while the X-ray bolometric ratio decreases with a power law behaviour for $R_0 > 0.13$ (non-saturated regime). 

Our measurements are fully consistent with the \citet{Wright2018} relation and -- more importantly -- with its large scatter, which means that inferring the stellar $L_{X}$ on the basis of properties such as rotation and convection alone can lead to exceedingly large errors (up to two orders of magnitude) in the resulting atmospheric mass loss rates. In addition, a non negligible fraction of our sample stars (HD 97658, HD 189733, HAT-P-11, 55 Cnc, GJ 436, Au Mic) exhibit significant X-ray variability (up to 225\%), which can further impact the resulting mass outflows (up to 135\%). 
Integrated over the system lifetime, these uncertainties could make the difference between retaining vs. losing the atmosphere. \\

In the following, we discuss our revised X-ray luminosities vis a vis previous values reported in the literature, when available. 
For the purpose of a systematic comparison with thee mass loss rates, whenever possible we refer to the work by \citet{Salz2016a} (hereafter S16), and references therein, who assembled X-ray measurements/estimates for most of our targets, i.e.: HAT-P-2, WASP-18, HAT-P-20, WASP-10, WASP-38, WASP-8, WASP-43, WASP-77 A, HD 189733, WASP-80, HD 209458, HD 149026, HAT-P-11, 55 Cnc, HD 97658, GJ 436, GJ 3470, GJ 1214. In addition, our sample includes WASP-29, WASP-107, HD 219134, GJ 9827, K2-25, LHS 1140, AU Mic; when available, we compare our results for these systems, too, with the literature.

\subsection{HAT-P-2}

Our re-analysis of the Chandra observation yields a flux of 5.50$^{+0.74}_{-0.79}$ $\times$ 10$^{-14}$ erg cm$^{-2}$ s$^{-1}$, in agreement with the value obtained by \citet{Salz2015} and reported by S16. The revised GDR2 distance (127.77 $\pm$ 0.42 pc vs. 114 $\pm$ 10 pc) implies a luminosity of 1.07 $\pm$ 0.15 $\times$ 10$^{29}$ erg s$^{-1}$. 

\subsection{WASP-18}
WASP-18 is not significantly detected by Chandra. We estimate a 3$\sigma$ upper limit of 0.10 $\times$ 10$^{-14}$ erg cm$^{-2}$ s$^{-1}$, higher than that found by \citet{Salz2015} and reported by S16 (0.06 $\times$ 10$^{-14}$ erg cm$^{-2}$ s$^{-1}$). The revised GDR2 distance (123.48 $\pm$ 0.37 pc vs. 99 $\pm$ 10 pc) implies a higher yet X-ray luminosity upper limit, of 1.82 $\times$ 10$^{27}$ erg s$^{-1}$ (vs. 7 $\times$ 10$^{26}$ erg s$^{-1}$ in S16).

\subsection{HAT-P-20}

Based on our re-analysis of the Chandra observation we measure a flux of 1.78$^{+0.28}_{-0.35}$ $\times$ 10$^{-14}$ erg cm$^{-2}$ s$^{-1}$, in excellent agreement with the value reported by \citet{Salz2015} and used by S16. At a distance of 71.03 pc (GDR2) this measurement implies an X-ray luminosity of 1.07$^{+0.17}_{-0.21}$ $\times$ 10$^{28}$ erg s$^{-1}$.

\subsection{WASP-10}

The target is only marginally detected in the Chandra data, at the $1.7\sigma$ confidence level. Assuming an APEC model with temperature 0.3 keV, we derive a $3\sigma$ upper limit of $<$ 0.90 $\times$ 10$^{-14}$ erg cm$^{-2}$ s$^{-1}$, slightly more stringent than the value of 1.3$^{+0.01}_{-1.3}$ $\times$ 10$^{-14}$  erg cm$^{-2}$ s$^{-1}$ reported by S16. Adopting the new GDR2 distance (141.00 $\pm$ 0.75 pc vs. 90 $\pm$ 20 pc), the X-ray luminosity (limit) for this system increases by a factor of 2.5, to $\simlt$ 2.14 $\times$ 10$^{28}$ erg s$^{-1}$.

\subsection{WASP-38}
For this system, the Chandra data yield a $3\sigma$ limit of 0.56 $\times$ 10$^{-14}$ erg cm$^{-2}$ s$^{-1}$ (vs. 0.8 $\times$ 10$^{-14}$ erg cm$^{-2}$ s$^{-1}$ given by S16). 
Considering the new GDR2 distance (136.24 $\pm$ 0.80 pc vs. 110 $\pm$ 20 pc) the resulting X-ray luminosity (limit) is $\simlt$ 1.24 $\times$ 10$^{28}$ erg s$^{-1}$.

\subsection{WASP-8}
Our re-analysis of the Chandra data yields a flux of 3.50$^{+0.61}_{-0.71}$ $\times$ 10$^{-14}$ erg cm$^{-2}$ s$^{-1}$, consistent within the errors with the value reported by S16. At a distance of 89.69 pc (GAIA DR2) this measurement implies an X-ray luminosity of 3.37$^{+0.32}_{-1.01}$ $\times$ 10$^{28}$ erg s$^{-1}$.

\subsection{WASP-43}
Our re-analysis of the XMM observation of this system yields a flux of 0.75$^{+0.12}_{-0.20}$ $\times$ 10$^{-14}$ erg cm$^{-2}$ s$^{-1}$, consistent within the errors with the value listed by S16, and derived by \citet{Czesla2013}. Adopting the new GDR2 distance (86.75 $\pm$ 0.33 pc vs 80 $\pm$ 30 pc) this measurement implies an X-ray luminosity of 6.75$^{+1.09}_{-1.80}$ $\times$ 10$^{27}$ erg s$^{-1}$, still consistent within the errors with the value reported by S16.

\subsection{WASP-77 A}
Based on the archival Chandra observation we measure a flux of 1.24$^{+0.16}_{-0.35}$ $\times$ 10$^{-14}$  erg cm$^{-2}$ s$^{-1}$ in excellent agreement with the value listed S16, and derived by \citet{Salz2015}.
Adopting the new GDR2 distance (105.17 $\pm$ 1.20 pc vs 93 $\pm$ 5 pc) this measurement implies an X-ray luminosity of 1.64$^{+0.21}_{-0.46}$ $\times$ 10$^{28}$  erg s$^{-1}$, consistent within the errors with the value reported by S16.

\subsection{WASP-80}

This system was observed twice with XMM; we measure a flux of [1.77$^{+0.16}_{-0.28}$--1.70$^{+0.11}_{-0.19}$] $\times$ 10$^{-14}$ erg cm$^{-2}$ s$^{-1}$ for observations A and B respectively, consistent within the errors with the value quoted by S16, and based again on \citet{Salz2015}, as well as \cite{King2018} (K18 hereafter). The revised GDR2 distance (49.78 $\pm$ 0.12 pc vs. 60 $\pm$ 20 pc) implies a luminosity between [5.26$^{+0.47}_{-0.84}$--5.05$^{+0.32}_{-0.57}$] $\times$ 10$^{27}$ erg s$^{-1}$.

\subsection{HD 209458} 
                                                                            
This system was observed four times, three with XMM (although observation B was affected by a filter-wheel failure) and once with Chandra, but only detected in the Chandra data (observation D). For that, we measure a flux of 2.25$^{+1.12}_{-2.25}$  $\times$ 10$^{-14}$ erg cm$^{-2}$ s$^{-1}$, where the large uncertainties are due to the degeneracy between the emission measure and the temperature in the model. The most stringent $3\sigma$ upper limit measured by XMM (observation C) corresponds to a flux $\simlt$ $0.26$ $\times$ 10$^{-14}$ erg cm$^{-2}$ s$^{-1}$, indicating statistically significant X-ray variability. This value is less stringent than the value obtained by \citet{Sanz-Forcada2011} (and reported by S16) with the same data ($\simlt$ 0.10 $\times$ 10$^{-14}$ erg cm$^{-2}$ s$^{-1}$). At a distance of 48.30 pc (GDR2), these measurements imply an X-ray luminosity range between $\simlt$ 7.26 $\times$ $10^{26}$ erg s$^{-1}$ and  6.28$^{+3.13}_{-6.28}$ $\times$ $10^{27}$ erg s$^{-1}$.

\subsection{HD 149026}

Based on the archival XMM observation we measure a flux of 0.75$^{+0.04}_{-0.21}$ $\times$ 10$^{-14}$ erg cm$^{-2}$ s$^{-1}$, in agreement with the value reported by K18 based on the same data. S16 reported a significantly higher flux (5.33 $\times$ 10$^{-14}$ erg cm$^{-2}$ s$^{-1}$) on the basis of the empirical relation with stellar mass and rotation rate derived by \citet{Pizzolato2003} for a sample of main sequence stars. As suggested by K18, the discrepancy is likely due to the sub-giant nature of HD 149026. At a distance of 79.86 pc (GDR2) the resulting X-ray luminosity of 5.72$^{+0.31}_{-1.60}$ $\times$ $10^{27}$ erg s$^{-1}$ is a factor 7 lower than the value reported by S16. 

\subsection{WASP-29}

WASP-29 is not significantly detected by XMM. We derive a $3\sigma$ upper limit of 0.47 $\times$ 10$^{-14}$ erg cm$^{-2}$ s$^{-1}$. Based on the same observation \citet{DosSantos2021} found a more stringent upper limit, of 3.68 $\times$ 10$^{-16}$ erg cm$^{-2}$ s$^{-1}$. At a distance of 87.60 pc (GDR2) we obtain an X-ray luminosity (limit) of $\simlt$ 4.32 $\times$ 10$^{27}$ erg s$^{-1}$.

\subsection{WASP-69}
Based on the archival XMM observation we measure a flux of 6.02$^{+0.27}_{-0.20}$ $\times$ 10$^{-14}$ erg cm$^{-2}$ s$^{-1}$, higher than the flux measured by \citet{Nortmann2018}  (4.79$^{+0.15}_{-0.16}$ $\times$ 10$^{-14}$ erg cm$^{-2}$ s$^{-1}$).  At a distance of 49.96 pc (GDR2) the resulting X-ray luminosity is 1.80$^{+0.08}_{-0.07}$ $\times$ $10^{27}$ erg s$^{-1}$.

\subsection{HAT-P-11}

This system was observed once with XMM and once with Chandra. For the former data set, we measure a flux of 3.25$^{+0.17}_{-0.28}$ $\times$ 10$^{-14}$ erg cm$^{-2}$ s$^{-1}$, consistent with the value derived by K18, as well as (within the errors) with that reported by S16 on the basis of the \citet{Pizzolato2003} relation.  We also report, for the first time, on the Chandra observation (observation B in Table~\ref{tab:x2}), for which we measure a flux of 7.84$^{+0.84}_{-0.72}$ $\times$ 10$^{-14}$ erg cm$^{-2}$ s$^{-1}$, implying statistically significant variability over a 6 month timescale.
At a distance of 37.77 pc, these measurements imply an X-ray luminosity in the range [5.55$^{+0.33}_{-0.49}$--13.38$^{+1.44}_{-1.23}$] $\times$ 10$^{27}$ erg s$^{-1}$.

\subsection{55 Cnc} 
This system was observed once with XMM and twice with Chandra (obs. A, B and C, respectively). For observation A, we derive a flux of 8.65$^{+0.71}_{-1.96}$ $\times$ 10$^{-14}$ erg cm$^{-2}$ s$^{-1}$, in agreement within the errors with the analysis of \citet{Poppenhaeger2010}, and a factor about 3.6 times higher than estimated by \citet{Sanz-Forcada2011} and, given by S16 (2.4 $\times$ 10$^{-14}$ erg cm$^{-2}$ s$^{-1}$). The two Chandra observations, which we report on for the first time, yield a $3 \sigma$ upper limit of $\simlt$1.89 $\times$ 10$^{-14}$ erg cm$^{-2}$ s$^{-1}$ for observation B and a detection of 0.77$^{+0.32}_{-0.19}$ $\times$ 10$^{-14}$ erg cm$^{-2}$ s$^{-1}$ for observation C, implying significant variability. At a distance of 12.59 pc (GDR2), this implies an X-ray luminosity in the range [$\simlt$3.59  s$^{-1}$--1.46$^{+0.61}_{-0.36}$--16.41$^{+1.41}_{-3.70}$] $\times$ 10$^{26}$ erg s$^{-1}$. 

\subsection{HD 97658}

This target was observed once with XMM (observation A in Table \ref{tab:x2}) and three times with Chandra (B, C and D). For the XMM observation we measure an X-ray flux of 3.20$^{+0.06}_{-0.78}$ $\times$ 10$^{-14}$ erg s$^{-1}$ cm$^{-2}$, consistent within the errors with the analysis of \citet{Bourrier2017} and K18, as well as with the value quoted by S16 on the basis of the empirical relation between $L_{X}$, stellar rotation period, and stellar mass obtained by \citep{Pizzolato2003}. Analysis of the three Chandra observations confirms the presence of variability \citet{Bourrier2017}; we measure X-ray fluxes of [1.73$^{+0.61}_{-0.75}$--0.98$^{+0.35}_{-0.51}$--$\simlt$1.32] $\times$ 10$^{-14}$ erg cm$^{-2}$ s$^{-1}$ for the B/C/D observations, respectively. At a distance of 21.56 pc (GDR2), combining the XMM and Chandra results we derive an X-ray luminosity in the range [0.55$\pm$0.30--1.78$^{+0.03}_{-0.43}$] $\times$ 10$^{27}$ erg s$^{-1}$. 

\subsection{WASP-107}

Based on the archival XMM observation we measure a flux of 1.42$^{+0.01}_{-0.02}$ $\times$ 10$^{-14}$ erg cm$^{-2}$. This is 2.5 times lower than estimated by \citet{Nortmann2018} based on the same data. At a distance of 64.74 pc (GDR2) we derive an X-ray luminosity of 7.12$\pm$0.05 $\times$ 10$^{27}$ erg s$^{-1}$.

\subsection{HD 219134}

Based on the archival XMM observation, which we analyze for the first time, we obtain a flux of 18.07$^{+0.40}_{-0.89}$ $\times$ 10$^{-14}$ erg cm$^{-2}$. At a distance of 6.53 pc (GDR2) we derive an X-ray luminosity of  9.22$^{+0.20}_{-0.46}$ $\times$ 10$^{26}$ erg s$^{-1}$.

\subsection{GJ 9827}

This system was observed three times with XMM (A,B,C). We analyze these observations for the first time, yielding a flux of [0.68$^{+0.13}_{-0.12}$--0.65$^{+0.10}_{-0.15}$--0.43$^{+0.15}_{-0.09}$] for observation A/B/C, respectively. At a distance of 29.66 pc, these measurements imply an X-ray luminosity in the range [$4.53^{+1.58}_{-0.95}$--$7.16^{+1.37}_{-2.64}$] $\times$ 10$^{26}$ erg s$^{-1}$.

\subsection{K2-25}
Based on the archival XMM observation we obtain a flux of 8.61$^{+0.50}_{-0.45}$ $\times$ 10$^{-14}$ erg cm$^{-2}$, consistent within the errors with the value reported in \citet{Rockcliffe2021} based on the same observation. At a distance of 44.96 pc (GDR2) we derive an X-ray luminosity of 2.08$^{+0.13}_{-0.12}$ $\times$ 10$^{28}$ erg s$^{-1}$. 

\subsection{GJ 436}

This system has eight archival observations, two with XMM (obs A and H) and 6 with Chandra (obs. B-G). 
For the XMM data we derive fluxes of 2.37$^{+0.15}_{-0.54}$ and 2.93$^{+0.20}_{-0.33}$ $\times$ 10$^{-14}$ erg cm$^{-2}$ s$^{-1}$ respectively for obs. A and H, both in excellent agreement with the values reported by K18. However our obs. A flux is 4 times higher than that measured by \citet{Sanz-Forcada2011} (and given by S16) and 1.7 lower than that estimated by \citealt{Ehrenreich2015}. We also analyzed the 6 Chandra  observations, for which we estimate fluxes of [2.29$^{+0.26}_{-1.79}$--1.80$^{+0.33}_{-0.45}$--1.60$^{+0.22}_{-0.36}$--1.97$^{+0.37}_{-0.43}$--1.79$^{+0.34}_{-0.63}$--2.86$^{+0.53}_{-0.50}$] $\times$ 10$^{-14}$ erg cm$^{-2}$ s$^{-1}$ for B/C/D/E/F/G respectively. At a distance of 9.75 pc (GDR2) our fluxes imply an X-ray luminosity in the range [$1.82^{+0.25}_{-0.41}$--$3.33^{+0.23}_{-0.61}$] $\times$ 10$^{26}$ erg s$^{-1}$.

\subsection{LHS 1140}

We report on the analysis of the XMM observation of this target for the first time. We measure an X-ray flux of 0.50$^{+0.07}_{-0.08}$ $\times$ 10$^{-14}$ erg cm$^{-2}$ s$^{-1}$, consistent with the upper limit reported by \citet{Spinelli2019} based on a Swift telescope observation. At a distance of 14.99 pc (GDR2) we derive an X-ray luminosity of 1.34$^{+0.19}_{-0.21}$ $\times$ 10$^{26}$ erg s$^{-1}$.

\subsection{AU Mic}

AU Mic was observed five times with XMM (obs A-E). We measure fluxes between [2209 $\pm$ 4 -- 3164$^{+4}_{-5}$] $\times$ 10$^{-14}$ erg cm$^{-2}$ s$^{-1}$. At a distance of 9.72 pc (GDR2) these imply an X-ray luminosity in the range [2.498$^{+0.004}_{-0.005}$--3.577$^{+0.005}_{-0.006}$] $\times$ 10$^{29}$ erg s$^{-1}$.

\subsection{GJ 3470}

Our re-analysis of the XMM observation yields a flux of 3.68$^{+0.31}_{-0.25}$ $\times$ 10$^{-14}$ erg cm$^{-2}$ s$^{-1}$, consistent within the errors with the value measured by K18 based on the same observation. Based of the \citet{Pizzolato2003} relation, S16 give 4.24 $\times$ 10$^{-14}$ erg cm$^{-2}$ s$^{-1}$. At a distance of 29.42 (GDR2) our flux implies an X-ray luminosity of 3.81$^{+0.32}_{-0.26}$ $\times$ 10$^{27}$ erg s$^{-1}$.

\subsection{GJ 1214}

This system was observed once with XMM (obs A) and once with Chandra (obs B). For the XMM data, we derive a flux of 0.62$^{+0.27}_{-0.05}$ $\times$ 10$^{-14}$ erg cm$^{-2}$ s$^{-1}$, formally inconsistent with the value reported by 
\citet{Lalitha2014} and used by S16 (the difference arise in the choice of the best-fit model; we use a two temperature APEC model with coronal temperatures free to vary between 0.01-2.00 keV, whereas \citealt{Lalitha2014} assumed a single temperature APEC model). For the Chandra observation we found an upper limit of $\simlt$0.32 $\times$ 10$^{-14}$ erg cm$^{-2}$ s$^{-1}$. At a distance of 16.79 pc (GDR2), this translates into a luminosity range of $\simlt$1.08--2.09$^{+0.91}_{-0.17}$ $\times$ 10$^{26}$ erg s$^{-1}$.

\subsection{HD 189733}
\label{sec:hd189733}

Because of the large number of pointings available, the observations for this system are listed in a separate table, i.e. Table~\ref{tab:x3}.  The flux quoted by S16 refers to the work of \citet{Sanz-Forcada2011}, based on the first observation of the system, with XMM. 
For the same observation (A), we measure a flux of 3.35$\pm$ 0.07 $\times$ 10$^{-13}$ erg cm$^{-2}$ s$^{-1}$, in excellent agreement with \citet{Sanz-Forcada2011}. In addition, we perform a re-analysis of 30 other archival observations -- 24 more with XMM-Newton, and 6 with Chandra -- taken between 2007 and 2015.  We confirm that the system exhibits statistically significant X-ray variability, between [2.78$^{+0.08}_{-0.10}$--5.68$^{+0.06}_{-0.07}$] $\times$ 10$^{-13}$ erg cm$^{-2}$ s$^{-1}$, in broad in agreement with \citet{Pillitteri2022} (XMM) and \citet{Poppenhaeger2010} (Chandra). At a distance of 19.76 pc (GDR2), this yields a luminosity range between [1.30$^{+0.04}_{-0.05}$--2.65$\pm$0.03] $\times$ 10$^{28}$ erg s$^{-1}$.

\begin{figure}
    \center
	\includegraphics[width=1.0\columnwidth]{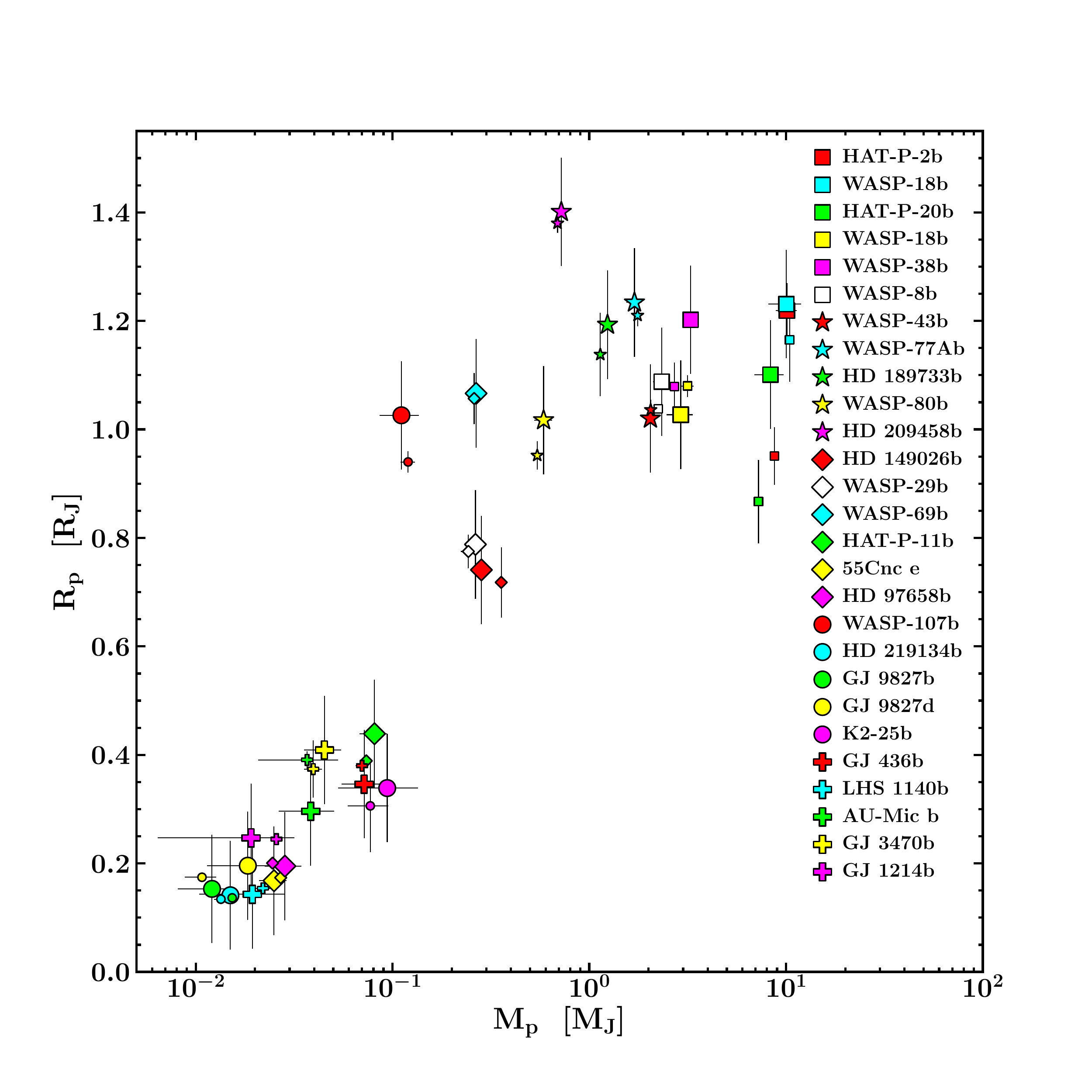}
	\caption{Revised masses and radii (in Jovian units) for the planet sample under consideration are shown as large symbols, and compared against the values reported in the \protect\hyperlink{exoplanet.eu}{exoplanet.eu} catalog, shown as smaller symbols.}
	\label{fig:massradius}
\end{figure}

\begin{table*}
	\caption{Planetary parameters after GDR2 distances; stellar density (1), stellar mass (2), planet  radius (3), planet mass (4), average orbital distance (5), EUV irradiation (i.e., flux at the planet) (6), ATES mass loss rates (7).}
\label{tab:mdot}
\fontsize{7}{13}\selectfont
\renewcommand{\arraystretch}{1.0}
\small
\begin{tabular}{
ccccccccc}
    	\hline
    	\hline
    	& %
    	\boldmath $\rho_{\star}$ & 
            \boldmath $M_{\star}$ & 
            \boldmath$R_{p}$ &
            \boldmath$M_{p}$ &
            \boldmath$a$ &
             \boldmath$\log(F_{\rm EUV})$ &
             \boldmath $\log(\dot{M})$ \\

            &
      $\mathbf{[\rho_{\odot}]}$     & %
    	 $\mathbf{[M_{\odot}]}$           & %
    	  $\mathbf{[R_{J}]}$ 			        & %
    	  $\mathbf{[M_{J}]}$ 				    & %
    	   $\mathbf{[AU]}$						    & 
        $\mathbf{[erg ~ s^{-1} cm^{-2}]}$                      & %
           $\mathbf{[g ~ s^{-1}]}$ 	  \\
            & 
        (1) & 
        (2) & 
        (3) & 
        (4) & 
        (5) & 
        (6) &
        (7) \\
    	\cline{1-8} 
    	Host FGK & & & & & & &  \\
    	\cline{1-8}
        HAT-P-2   b & 0.31$\pm$0.04  & 1.60$\pm$0.29 & 1.22$\pm$0.05 & 10.11$\pm$1.23 & 0.082$\pm$0.005 & 4.02$^{+0.03}_{-0.04}$     &\textbf{-}  \\
        WASP-18   b & 0.64$\pm$0.09  & 1.22$\pm$0.34 & 1.23$\pm$0.10 & 10.03$\pm$1.90  & 0.020$\pm$0.002     & -       &\textbf{-}   \\ 
        HAT-P-20  b & 2.38$\pm$0.16  & 0.93$\pm$0.23 & 1.10$\pm$0.12 & 8.31$\pm$1.42  & 0.039$\pm$0.003  & 3.80$^{+0.03}_{-0.05}$      &\textbf{-}  \\
        WASP-10   b & 2.16$\pm$0.08  & 0.65$\pm$0.15 & 1.03$\pm$0.08 & 2.91$\pm$0.44  & 0.036$\pm$0.003 &  -       & \textbf{-}   \\
        WASP-38   b & 0.51$\pm$0.02  & 1.62$\pm$0.21 & 1.20$\pm$0.16 & 3.27$\pm$0.29  & 0.083$\pm$0.004 &     &  -      &\textbf{-}  \\
        WASP-8    b & 1.22$\pm$0.16  & 1.18$\pm$0.17 & 1.09$\pm$0.03 & 2.33$\pm$0.23  & 0.088$\pm$0.004 &       3.47$^{+0.04}_{-0.05}$      &  \textbf{-} \\
        WASP-43   b & 2.49$\pm$0.21  & 0.72$\pm$0.20 & 1.02$\pm$0.10 & 2.04$\pm$0.39  & 0.015$\pm$0.001 &      4.49$^{+0.03}_{-0.07}$      & \textbf{7.97} \\
        WASP-77 A b & 1.10$\pm$0.04  & 0.94$\pm$0.12 & 1.24$\pm$0.05 & 1.70$\pm$0.15 & 0.024$\pm$0.001 & 4.43$^{+0.03}_{-0.08}$        & \textbf{9.60}  \\
        HD 189733 b & 1.98$\pm$0.17  & 0.94$\pm$0.11 & 1.19$\pm$0.06 & 1.24$\pm$0.10  & 0.033$\pm$0.001  & 4.02$^{+0.01}_{-0.01}$-4.19$^{+0.01}_{-0.01}$   & \textbf{9.78-9.99}  \\
        WASP-80 b    & 2.87$\pm$0.01  & 0.65$\pm$0.10 & 1.02$\pm$0.05 & 0.59$\pm$0.06 & 0.036$\pm$0.002 &  3.65$^{+0.02}_{-0.04}$      &\textbf{10.57}  \\
        HD 209458 b  & 0.73$\pm$0.03  & 1.20$\pm$0.05 & 1.40$\pm$0.01 & 0.72$\pm$0.02 & 0.048$\pm$0.001  & 3.76      & \textbf{11.12}   \\
        HD 149026 b  & 0.34$\pm$0.03  & 1.09$\pm$0.13 & 0.74$\pm$0.02 & 0.28$\pm$0.03 & 0.041$\pm$0.002     & 3.93$^{+0.01}_{-0.07}$      &\textbf{11.08}  \\
        WASP-29 b	 & 1.70$\pm$0.02  & 0.92$\pm$0.10 & 0.79$\pm$0.03 & 0.26$\pm$0.03 &  0.047$\pm$0.002  & -       & \textbf{-} \\
        WASP-69 b	 & 1.55$\pm$0.18 & 0.85$\pm$0.14 & 1.07$\pm$0.04 & 0.26$\pm$0.03 & 0.046$\pm$0.002      & 3.85$^{+0.01}_{-0.01}$      & \textbf{11.50}  \\
        HAT-P-11 b   & 1.76$\pm$0.04  & 0.80$\pm$0.16 & 0.44$\pm$0.03 & 0.081$\pm$0.013 & 0.054$\pm$0.003 & 3.41$^{+0.02}_{-0.02}$-3.57$^{-0.02}_{+0.02}$  & \textbf{10.41-10.59}  \\
        55 Cnc e     & 1.08$\pm$0.01  & 0.93$\pm$0.20 & 0.17$\pm$0.01 & 0.026$\pm$0.004 & 0.016$\pm$0.001 & 3.85$^{+0.08}_{-0.07}$ -4.32$^{+0.08}_{-0.13}$   & \textbf{10.22-10.59} \\
        HD 97658 b	 & 2.16$\pm$0.50  & 0.91$\pm$0.24 & 0.20$\pm$0.01 & 0.028$\pm$0.006 & 0.085$\pm$0.007 &  2.53$^{+0.07}_{-0.17}$-2.76$^{-0.03}_{+0.03}$   &\textbf{9.01-9.24} \\
        WASP-107 b	 & 2.17$\pm$0.01  & 0.85$\pm$0.28 & 1.03$\pm$0.11 & 0.11$\pm$0.03 & 0.059$\pm$0.006  & 3.37$^{+0.01}_{-0.01}$       & \textbf{11.35} \\
        HD 219134 b	 & 1.62$\pm$0.05  & 0.84$\pm$0.10 & 0.14$\pm$0.03 & 0.015$\pm$0.001 &  0.039$\pm$0.002 & 3.32$^{+0.01}_{-0.01}$     & \textbf{9.61} \\
        GJ 9827   b  & 2.67$\pm$0.45  & 0.73$\pm$0.24 & 0.15$\pm$0.01 & 0.012$\pm$0.004 & 0.020$\pm$0.002 &  3.76$^{+0.04}_{-0.04}$        &  \textbf{10.17} \\
        GJ 9827   d	 & 2.67$\pm$0.45  & 0.73$\pm$0.24 & 0.20$\pm$0.02 & 0.018$\pm$0.007 & 0.060$\pm$0.006  &  2.81$^{+0.04}_{-0.04}$        & \textbf{9.42} \\
    	\cline{1-8}
        Host M & & & & & & &  \\
    	\cline{1-8}
    	K2-25 b        & 10.37$\pm$0.86 & 0.35$\pm$0.20  & 0.34$\pm$0.07 & 0.094$\pm$0.041  & 0.034$\pm$0.007       & 3.72$^{+0.01}_{-0.01}$        & \textbf{10.44}    \\
        GJ 436 b       & 5.90$\pm$0.18  & 0.47$\pm$0.16  & 0.35$\pm$0.04 & 0.072$\pm$0.017  & 0.029$\pm$0.003 & 2.97$^{+0.03}_{-0.06}$-3.08$^{+0.04}_{-0.04}$   & \textbf{9.71-9.79}   \\
        LHS 1140 b	   & 19.65$\pm$1.34 & 0.18$\pm$0.13  & 0.14$\pm$0.03& 0.019$\pm$0.013  & 0.094$\pm$0.022 & 1.56$^{+0.04}_{-0.03}$        & \textbf{7.72}       \\
        AU Mic b	   & 1.31$\pm$0.06  & 0.30$\pm$0.08  & 0.30$\pm$0.02 & 0.038$\pm$0.013  & 0.055$\pm$0.005 & 4.14$^{+0.01}_{-0.01}$-4.22$^{+0.01}_{-0.01}$   & \textbf{11.18-11.29} \\
        GJ 3470 b      & 3.27$\pm$0.33  & 0.54$\pm$0.16  & 0.41$\pm$0.04 & 0.045$\pm$0.010  & 0.036$\pm$0.003 & 3.56$^{+0.02}_{-0.03}$        & \textbf{10.65}      \\
        GJ 1214	b      & 16.79$\pm$1.03 & 0.17$\pm$0.14  & 0.25$\pm$0.08 & 0.019$\pm$0.013  & 0.015$\pm$0.005      & 3.24$^{+0.08}_{-0.02}$        & \textbf{10.09}      \\
    	\hline
    	\hline
	\end{tabular}
\end{table*}

\section{Planetary parameters}
\label{sec:param}
Reliable estimates of planetary parameters require accurate measurements of the host star properties. In this section we describe how we use the parallactic distances from GDR2  
to derive revised estimates of planetary radii $R_p$, masses $M_p$, and orbital separations $a$. As noted above, all systems under consideration have both transit and radial velocity measurements, and thus known orbital inclinations. 

For circular orbits, the stellar density $\rho_\star$ can be obtained directly from the photometric transit data, namely the total transit duration, $T$; the duration of the ``flat part" of the transit, $t_F$ (between the so-called second and the third contacts); the orbital period, $P$; and the transit depth $\Delta F$. These yield the ratio between the orbital semi-major axis $a$ and the stellar radius $R_\star$ \citep{Seager2003}.

As shown by \citet{Kipping2010, Moorhead2011, Tingley2011, Dawson2012, VanEylen2015}, this approach can be generalized to non circular orbits with known eccentricity and periastron argument ($\omega$), e.g., from secondary transit timing, transit timing variations, or radial velocity curves. In this case we have \citep{Kipping2010}:
 \begin{equation}
   \rho_{\star} \simeq \frac{3 \pi}{G P^2} \left(\frac{a}{R_{\star}}\right)^3\,  \frac{(1-e^2)^{3/2}}{(1+e \sin{\omega})^3} =  \frac{3 \pi}{G P^2} \left(\frac{a}{R_{\star}}\right)_{E}^3,
 \end{equation}
where $(a/R_{\star})_{E}$ denotes the eccentricity-corrected orbital separation to stellar radius ratio.\\

The next step consists of breaking the degeneracy between $R_\star$ and $M_\star$. A first approach proposed by, e.g., \citet{Hellier2011, Southworth2010, Bakos2011, McArthur2004, Maxted2013}, leverages stellar evolutionary models and combines distance-independent stellar parameters (such as effective temperature and metallicity) with the stellar density that is inferred from transit data to yield an estimate of mass and radius (see, e.g., \citealt{Rodriguez2021} for a discussion of the uncertainties at play in this method).
A second approach makes use of empirical mass-luminosity relations that are calibrated using dynamical mass measurements of stellar binaries \citep[e.g.,][]{Butler2004, Bonfilis2012}, under the assumption that the main stellar properties are not altered by the presence of a companion.  
A third approach, which we follow in this Paper, consists of estimating $R_\star$ of a star with known bolometric flux, effective temperature and distance, from Stefan-Boltzmann's law \citep{Stassun2017}. \\

For the FGK stars in our sample, we make use of the radius estimates derived by \citet{Andrae2018} based on GDR2 distances. For M stars, we apply the new luminosity–temperature–radius relations that were recently derived \citet{Morrell2019} accounting for the well-known radius inflation (compared to theoretical models) in such stars \\ \citep[e.g.][]{Birkby2012,Stassun2012, Mann2015, Somers2017}. 

Once the stellar radius and mass are known, the depth of the transit gives the (optical) planetary radius $R_p$. Finally, if $M_p \ll M_\star$, $M_p$ can be derived from the radial velocity semi-amplitude $K$ \citep{Torres2008}, as:
\begin{equation}
\small
    \frac{M_p \sin{i}}{M_{J}} = 4.919 \times 10^{-3} \left(\frac{K}{\rm m/s}\right) (1-e^2)^{1/2} \left(\frac{P}{\rm days}\right)^{1/3} \left(\frac{M_{\star}}{M_{\odot}}\right)^{2/3}, 
\end{equation}
where $i$ is the known orbital inclination angle with respect to the line of sight. \\

\begin{figure} 
    \center
	\includegraphics[width=1.0
	\columnwidth]{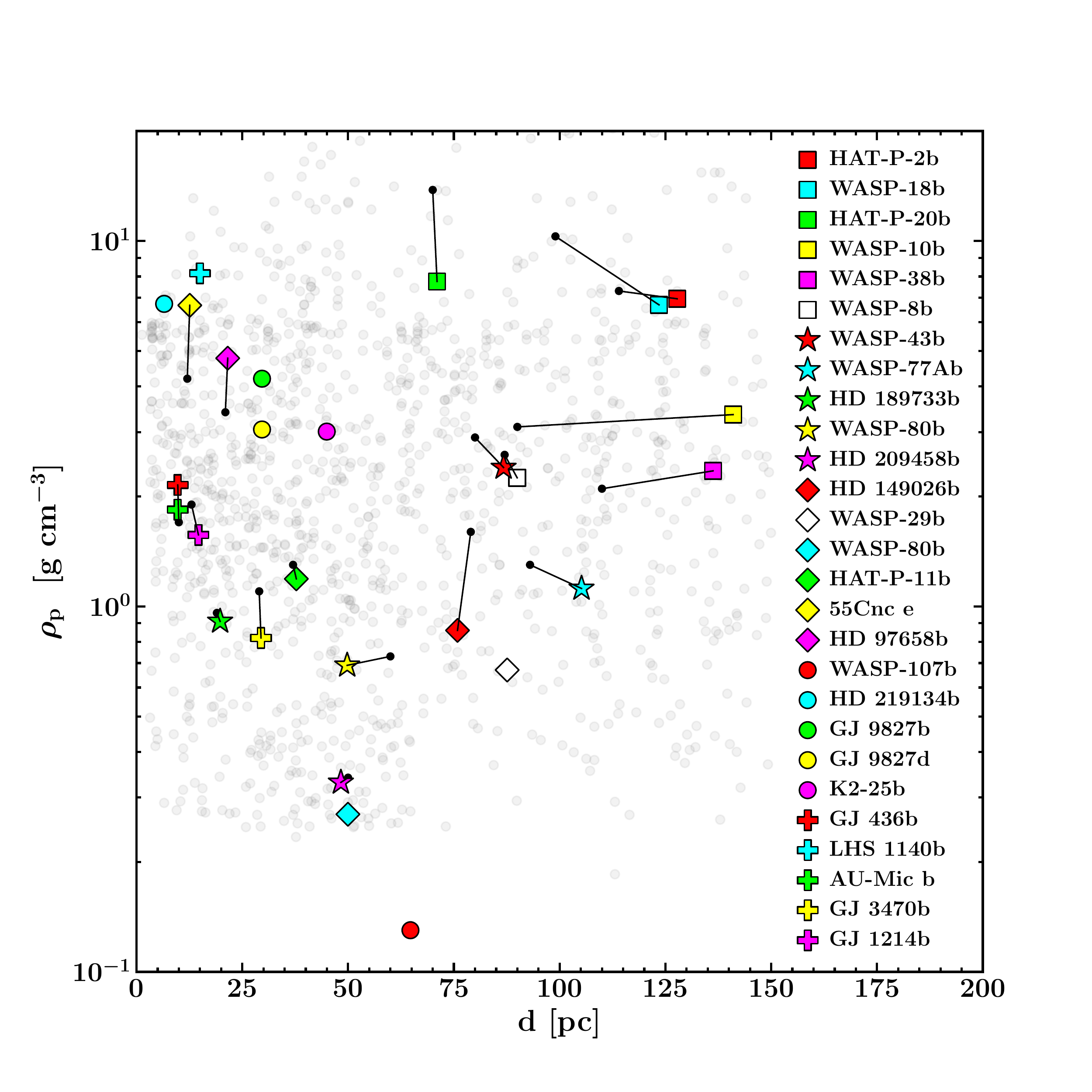}  
	\caption{Revised densities for the planet sample under consideration as a function of the Gaia DR2 distances. When applicable, these are compared to the values adopted by S16 (black dots). The grey circles in the background illustrate all the known planets within 150 pc.}
	\label{fig:distancedensity}
\end{figure}

The revised planetary parameters for our sample are listed in Table \ref{tab:mdot}; planetary mass and radii are shown in Figure \ref{fig:massradius}, and compared against the latest values listed on 
the \verb|exoplanet.eu|. Our method yields inconsistent values for the following five systems; for mass: HD 149026 b ($0.282\pm0.025$ $M_J$, vs. $0.357^{0.014}_{-0.011}$) and WASP-38 b ($3.272\pm 0.288$ $M_J$, vs. $2.712\pm 0.06$ ); for radius: Au Mic 20 ($0.296\pm0.0241$ $R_J$, vs. $0.3908\pm0.0161$), HAT-P-20 ($1.101\pm0.115$ $R_J$ vs $0.867\pm0.033$) and HAT-P-2 ($1.219\pm0.051$ $R_J$ vs. $0.951\pm0.053$). Figure \ref{fig:distancedensity} compares the resulting mass densities to those adopted by S16 for the purpose of estimating mass loss rates -- this is further explored next.

\section{EUV Irradiation and Mass loss rates}
\label{sec:mdot}

Mass loss rates are calculated using the 1D photoionization hydrodynamics code ATES\footnote{The code is publicly available at  \url{https://github.com/AndreaCaldiroli/ATES-Code}.} \citep{ates1}. ATES computes the temperature, density, velocity and ionization fraction profiles of a highly irradiated planetary atmosphere, and estimates the ensuing steady-state mass loss rate under the assumption of a primordial atmosphere (entirely composed of atomic hydrogen and helium).
As inputs we adopt the relevant parameters listed in Table~\ref{tab:planets_params} and~\ref{tab:mdot}, and specifically: stellar mass, planetary radius and mass, (eccentricity-corrected) average orbital distance, and planet equilibrium temperature.\\

Lacking a direct measurement of the stellar EUV flux  -- i.e., in the energy interval  [13.6, 124] eV -- we estimate the EUV irradiation experienced by the target planet(s) using the X-ray luminosities listed in Table~\ref{tab:flux}, and making use of the recently revised X-ray-to-EUV stellar correlations presented by \citet{King2018}. These authors perform a re-analysis of the same solar data that were used by \citet{Chadney2015} (see also \citealt{Woods2005}) to derive updated empirical relations between the surface EUV and X-ray fluxes in the energy bands where the Chandra and XMM-Newton X-ray telescopes are best calibrated (as opposed to the ROSAT observing band provided by Chadney et al.). Specifically, we adopt the relations of table 4 in \cite{King2018}.

Next, we convert the inferred stellar photo-ionizing fluxes into irradiation at the planet orbital separation. For planets with zero eccentricity (or with eccentricity upper limits) we adopt the semi-major axis reported in Table \ref{tab:mdot}; for non-zero eccentricity planets we use the time-averaged orbital separation defined by \cite{Williams2003}. In the case of K2-25 b ($e = 0.43$) and HAT-P-2 b ($e = 0.52$) the time-averaged separation is larger than the semi-major axis value by about 9\% and 13\%, respectively; this implies a decrease in the time-averaged flux experienced by the planet of 16\% and 22\%, respectively. The resulting EUV irradiation values are listed in Table \ref{tab:mdot}.\\

Since we are interested in realistic mass loss rate estimates (more so than limits), we attempt to run ATES only for those (23 out of 27) planets whose host stars have a statistically significant X-ray detection (22 out of 26). In those cases where X-ray variability is seen, we report the corresponding minimum and maximum mass loss value.
As expected based on the convergence study discussed in \cite{ates1}, ATES fails to converge for the three highest-gravity planets in the sample, namely HAT-P-20 b, WASP-8 b and HAT-P-2\footnote{The converge study shows that ATES recovers stable, steady-state solutions for systems with $\log(-\Phi_p) \lesssim 12.9 + 0.17\log F_{\rm XUV}$, where $\Phi_p$ and $F_{\rm XUV}$ are the planet gravitational potential and stellar irradiance in cgs units.}. 
In summary, we derive revised/new mass loss rates estimates for a total of 20 planets. Caution is warranted, however, for those systems with inferred radii close to 1.5 $R_{\Earth}$ (or 0.134 $R_{J}$), i.e., the lower bound of our selection criteria. Whereas this value is likely a lower limit to the minimum size of sub-Neptune-sized systems with H/He rich envelopes \citep{Lopez2014}, systems smaller than 2 $R_{\Earth}$ (0.178 $R_{J}$) are unlikely to have \textit{sizable} H/He atmospheres; this applies to 55~Cnc~e, HD~219134~b, GJ~9827~b, LHS 1140 b, and GJ 1214 b. 

\begin{figure} 
    \center
	\includegraphics[width=1.0
	\columnwidth]{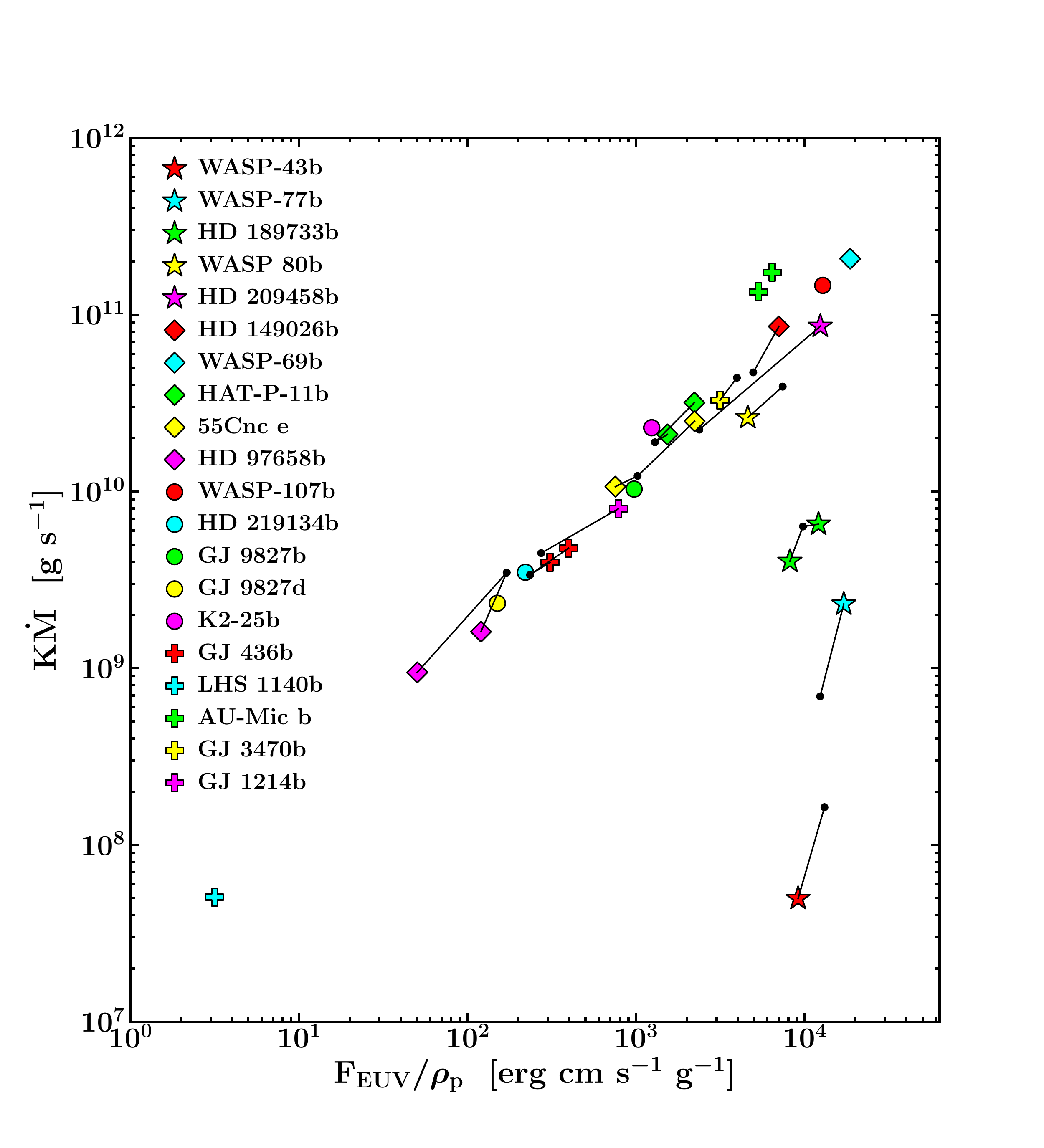}
	\caption{K-reduced mass outflow rates obtained with ATES, plotted as a function of the planetary density to irradiation ratio ($F_{\rm EUV}/\rho_p$) for the planet sample under consideration.The parameter $K$ accounts for the star contribution to the gravitational potential experienced by the atmosphere. 
	For those cases that were also considered by S16, we compare the outflow rates derived with the new (GDR2) planetary parameters 
	(color symbols) to those derived based on the ``old" values (black dots). Systems with two symbols represent those whose host star exhibits X-ray variability. }
	\label{fig:fsurho}
\end{figure}

Figure \ref{fig:fsurho} summarizes our results. Assuming that \textit{all} of the absorbed stellar X-ray and EUV flux is converted into expansion work, the instantaneous mass loss rate can be expected to scale linearly with the ratio $F_{\rm XUV}/\rho_p$,  with nearly 100\% efficiency \citep{Watson1981,Erkaev2007}. However, highly efficient hydrodynamic escape can only be attained below a specific planetary gravitational potential threshold (this was first pointed out by \citealt{Salz2016a}; see also \citealt{MurrayClay2009}).
Our results are fully consistent with with picture; the inferred mass loss rates scale (nearly) linearly with  $F_{\rm EUV}/\rho_p$, with the exception of three systems; WASP43 b, WASP77 b, and HD189733 b. Indeed, the gravitational potential energies of these planets exceed the threshold identified by \cite{atesII} ($\phi_{p}\simeq 14 \times 10^{12}$ erg g$^{-1}$). Above this value, the efficiency of hydrodynamic escape drops dramatically, regardless of irradiation. This is because, no matter how high $F_{\rm XUV}$ is, the mean kinetic energy acquired by the ions through photo-electron collisions in the atmosphere is insufficient for escape in the presence of such high planetary gravity.

\section{Summary and noteworthy results}\label{sec:summary}

In this paper, we make use of GAIA DR2 distances to deliver new or revised estimates of planetary parameters and X-ray irradiation for 27 gaseous planets (around 26 individual host stars) within 100 pc (Table \ref{tab:planets_params}), and with publicly available X-ray observations, either with Chandra or XMM-Newton (Table \ref{tab:x2}, \ref{tab:x3} and \ref{tab:flux}).
For 20 out of 22 planets with X-ray detected hosts (i.e., those within the convergence limits of the code) we derive updated atmospheric mass outflow rates making use of the 1D photoionization hydrodynamics code ATES (Table \ref{tab:mdot}). The planetary parameters derived in this work are adopted as inputs by \cite{atesII}, who consider a sub-sample of 16 out of the 27 planets analyzed here, and perform additional simulations with the purpose of exploring how the instantaneous mass loss rate is affected by varying $F_{\rm XUV}$. 
Below, we comment on specific systems for which the revised/new parameters and/or irradiation fluxes warrant attention. 

For five systems, the revised mass or radius do not agree with the latest values reported in the \verb|exoplanet.eu| archive. Worth noting is the  
resulting density for HD 149026 b, which has long been a matter of controversy. We infer a Saturn-like value of $0.86\pm 0.09$ g cm$^{-3}$, which is consistent with the lowest estimates reported in the literature. This independent estimate removes the need for a high metal fraction, which had been proposed to explain the allegedly high density of this system \citep{Sato2005, Charbonneau2006, Winn2008, Nutzman2009}. 

Separately, we report on the X-ray detection of 3 host stars for the first time: GJ 9827, HD 219134, and the M dwarf LHS 1140. The latter system is best known for LHS 1140~b: one of a handful of transiting planets within 15 pc to orbit within the habitable zone (HZ) of their star. 
Due to the shrunken HZs and higher stellar $L_{\rm XUV}/L_{\rm bol}$ ratios, planets around M dwarfs are expected to experience 10-200 times stronger XUV irradiation compared to those around FGK-type stars \citep{France2016, Becker2020}. This extreme environment may have negative implications for habitability \citep[e.g.][]{Shields2016}.

The X-ray detection of LHS 1140, with XMM, implies a luminosity of 1.34$^{+0.19}_{-0.21}$ $\times$ 10$^{26}$ erg s$^{-1}$, which is on the low-side for nearby M-dwarfs. Specifically, LHS 1140 is the 8th least X-ray luminous M-dwarf if compared to the sample of \citet{Stelzer2013}, who investigated the X-ray luminosity distribution of M dwarf stars within 10 pc (90\% completeness). The corresponding XUV irradiation upon LHS 1140~b is of 42 erg s$^{-1}$ cm$^{-2}$. For comparison, the 4 HZ planets around Trappist-1 \citep{Gillon2017} receive a XUV flux of $\sim$788, 457, 264 and 178 erg s$^{-1}$ cm$^{-2}$ \citep{Wheatley2016, Becker2020}. This makes LHS 1140 the least XUV-irradiated transiting HZ planet known around a nearby M dwarf, and thus a prime target for biosignature searches \citep{Wunderlich2021}.

\acknowledgments 
RS gratefully acknowledges partial support from the Michigan Institute for Research in Astrophysics. 

\bibliography{bibliography.bib}

\end{document}